\documentclass[a4paper]{article}
\usepackage[hmargin={2.3cm,2.3cm},vmargin={2.5cm,2.5cm},centering]{geometry}
\usepackage{amssymb,amsmath,amsthm,amsfonts,amscd}
\usepackage{mathabx}
\usepackage{graphicx,xcolor}
\usepackage{mathrsfs}
\usepackage[unicode,psdextra]{hyperref}
\usepackage{tikz-cd}
\usepackage[utf8]{inputenc}

\usepackage[english]{babel}
\usepackage{bbold}
\usepackage{csquotes}

\usepackage{mathtools}

\usepackage{enumerate}
\usepackage{setspace}

\begin{document}

\newtheorem{theorem}{Theorem}[section]
	
\newtheorem{defin}[theorem]{Definition}

\newtheorem{hyp}[theorem]{Hypothesis}

\newtheorem{lemma}[theorem]{Lemma}

\newtheorem{proposition}[theorem]{Proposition}

\newtheorem{cor}[theorem]{Corollary}

\newtheorem{remark}[theorem]{Remark}

\newtheorem*{teo*}{Theorem}
	
\newtheorem*{defin*}{Definition}

\newtheorem*{hyp*}{Hypothesis}

\newtheorem*{lem*}{Lemma}

\newtheorem*{propos*}{Proposition}

\newtheorem*{cor*}{Corollary}

\theoremstyle{remark}
	\newtheorem*{ex}{Example}

\theoremstyle{remark}
	\newtheorem{ass}{ASSUMPTION}

\newcommand{\RR}{\mathbb{R}}
\newcommand{\CC}{\mathbb{C}}
\newcommand{\NN}{\mathbb{N}}
\newcommand{\QQ}{\mathbb{Q}}
\newcommand{\ZZ}{\mathbb{Z}}
\newcommand{\real}{\Re\text{e}}

\newcommand{\der}{\partial}
\newcommand{\eps}{\varepsilon}
\newcommand{\HH}{\mathcal{H}}
\newcommand{\DD}{\mathcal{D}}
\newcommand{\scr}{\mathscr{S}}
\newcommand{\del}{\delta}

\newcommand{\la}{\langle}
\newcommand{\ra}{\rangle}
\newcommand{\ar}{\overrightarrow}

\newcommand{\Tr}{\mathrm{Tr}}
\newcommand{\tg}{\tan}
\newcommand{\supp}{\text{supp}}
\newcommand{\Ker}{\text{Ker}}
\newcommand{\rk}{\text{rk}}
\newcommand{\im}{\text{Im}}
\newcommand{\disp}{\displaystyle}
\newcommand{\tx}{\textstyle}

\newcommand{\one}{\mathds{1}}
\newcommand{\lf}{\left}
\newcommand{\ri}{\right}
\newcommand{\bra}[1]{\lf\langle #1\ri|}
\newcommand{\ket}[1]{\lf|#1 \ri\rangle}
\newcommand{\braket}[2]{\lf\langle #1\,|\,#2 \ri\rangle}
\newcommand{\braketr}[2]{\lf\langle #1\lf|#2\ri. \ri\rangle}
\newcommand{\braketl}[2]{\lf.\lf\langle #1\ri|#2 \ri\rangle}
\newcommand{\mean}[3]{\bra{#1}#2\ket{#3}}
\newcommand{\meanlr}[3]{\lf\langle #1\lf|#2\ri|#3\ri\rangle}

\newcommand{\RT}{\mathbb{R}^3}
\newcommand{\Rd}{\mathbb{R}^d}
\newcommand{\xv}{\mathbf{x}}
\newcommand{\xvp}{\mathbf{x}^{\prime}}
\newcommand{\yv}{\mathbf{y}}
\newcommand{\kv}{\mathbf{k}}
\newcommand{\kvp}{\mathbf{k}^{\prime}}
\newcommand{\gv}{\mathbf{g}}
\newcommand{\rv}{\mathbf{r}}
\newcommand{\sv}{\mathbf{s}}
\newcommand{\av}{\mathbf{a}}
\newcommand{\nv}{\mathbf{n}}
\newcommand{\mv}{\mathbf{m}}
\newcommand{\fv}{\mathbf{F}}

\newcommand{\diff}{\mathrm{d}}

\newcommand{\m}{\mathfrak{m}}
\newcommand{\eeps}{\epsilon}
\newcommand{\ceps}{c_{\eps}}
\newcommand{\dist}{\mathrm{dist}}
\newcommand{\gav}{\bm{\gamma}}
\newcommand{\nuv}{\bm{\nu}}
\newcommand{\tav}{\bm{\tau}}
\newcommand{\xiv}{\bm{\xi}}
\newcommand{\phiv}{\bm{\Phi}}
\newcommand{\ba}{\mathcal{B}}
\newcommand{\hamf}{H_{\mathrm{free}}}
\newcommand{\Z}{\mathbb{Z}}
\newcommand{\R}{\mathbb{R}}
\newcommand{\M}{\mathscr{M}}
\newcommand{\Mfin}{\mathscr{M}_{\mathrm{fin}}}
\newcommand{\N}{\mathbb{N}}
\newcommand{\kom}{\kappa}
\newcommand{\homega}{\mathfrak{H}_{\omega}}
\newcommand{\invomega}{\mathfrak{H}_{\omega^{-1}}}

\newcommand{\dom}{\mathscr{D}}
\newcommand{\F}{\mathcal{F}}

\newcommand{\A}{\mathcal{A}}
\newcommand{\B}{\mathcal{B}}
\newcommand{\T}{\mathcal{T}}

\newcommand{\HHe}{\mathcal{H}_{\mathrm{eff}}}
\newcommand{\KKe}{\mathcal{K}_{\mathrm{eff}}}

\newcommand{\FF}{\mathcal{F}}
\newcommand{\LL}{\mathcal{L}}
\newcommand{\Q}{\mathcal{Q}}
\newcommand{\OO}{\mathcal{O}}

\newcommand{\J}{\mathcal{J}}
\newcommand{\hh}{\mathfrak{H}}
\newcommand{\hilb}{\mathscr{H}}
\newcommand{\WW}{\mathcal{W}}
\newcommand{\esse}{\mathscr{S}}
\newcommand{\fock}{\Gamma_{\mathrm{sym}}}
\newcommand{\omegafield}{\diff\Gamma_{\eps}(\omega)}
\newcommand{\compactop}{\mathscr{L}_{\infty}(L^2)}
\newcommand{\traceclassop}{\mathscr{L}_{1}(L^2)}
\newcommand{\numberfield}{\diff\Gamma_{\eps}(\mathbb{1})}
\renewcommand{\emptyset}{\varnothing}
\renewcommand{\setminus}{\smallsetminus}
\renewcommand{\arraystretch}{1.5}

\title{The Casimir-Polder effect for an approximate Pauli-Fierz model: the atom plus wall case}
\author{M. Olivieri 	\\
\small Department of Mathematics, Aarhus University,  \small Ny Munkegade 118,  DK-8000 Aarhus C, Denmark\\
\small email: marco.olivieri@math.au.dk}
\date{}

\maketitle

\abstract{We study a system composed of a hydrogen atom interacting with an infinite conductor wall. The interaction energy decays like $L^{-3}$, where $L$ is the distance between the atom and the wall, due to the emergence of the van der Waals forces. In this paper we show how, considering the contributions from the quantum fluctuations of the electromagnetic field, the interaction is weakened to a decay of order $L^{-4}$, giving rise to the retardation effects which fall under the name of Casimir-Polder effect. The analysis is done by studying a suitable Pauli-Fierz model associated to the system, in dipole approximation and reduced to the interaction with $0$ and $1$ photons.}

\section*{Introduction} 

The intermolecular and interatomic interactions are at the basis of several important phenomena which occur in our world \cite{analewin, analewindiff, yu}. If we consider two neutral atoms, it is a well known fact \cite{pauling} that the fluctuations of the charge distribution of one atom create an instantaneous dipole which polarizes the other atom, and this triggers the emergence of multipole moments which influence back the dipole of the first atom. This process gives rise to an attractive interaction which is known as the van der Waals interaction. Van der Waals forces have universal decaying behaviors with respect to the distances between the interacting interfaces, and depend only on the geometry of the interfaces. There are two paradigmatic simple examples where this is evident: the interaction between two hydrogen atoms and the interaction between an hydrogen atom and an infinite, plane surface, perfect conductor (called, from now on, \enquote{wall}). Introducing the fine structure constant $\alpha$, whose approximate value is
\begin{equation}
\alpha \simeq \frac{1}{173} \ll 1,
\end{equation}
we work in suitable units of measurement such that $\hbar = c = 2m = 1$, where $\hbar, c, m$ are the reduced Planck constant, the speed of light and the electron mass, respectively, which implies that  
$\alpha$ is proportional to the square of electron charge: $e \simeq \sqrt{\alpha}$ and is equal to the inverse of the half of the Bohr radius: $a_0 = 2\alpha^{-1}$. We express the distance $y$ between the interfaces as multiple of Bohr radius, and therefore $\alpha^{-1}$, such that $y = L \alpha^{-1}$, for some $L \geq 1$.
If we denote by $W^{\text{qm}}_L$ the van der Waals energy at distance $L >0$ in a quantum mechanical description, \textit{i.e.}, where the interaction is considered instantaneous and originated only by the static Coulomb potentials, its decay for the aforementioned examples is
\begin{equation}\label{intro:decayQM}
W_L^{\text{qm}} \simeq 
\begin{dcases} 
-\frac{C_1(\alpha)}{L^6}, \qquad &\text{for two hydrogen atoms,}\\ 
-\frac{C_2(\alpha)}{L^3}, \qquad &\text{for a hydrogen atom and a wall},
\end{dcases}
\end{equation}
for suitable values of $C_1(\alpha),C_2(\alpha) >0$ in the chosen units system, see \cite{halfspace, anasigal,lennard}.
This description neglects, however, the retardation effects given by the interference with the quantum fluctuations of the field. If we take in consideration the fact that the electromagnetic field propagates at the speed of light (which is finite) the interaction is retarded. The behavior in \eqref{intro:decayQM} holds, indeed, up to a distance of approximately $10$ Bohr radii. At this distance, the information about the first atom's electron motion reaches the second interface in a time that is comparable with the average circulation time of the electron. This breaks the correlation between the two objects and weakens the interaction \cite{koppen}. This effect was studied and formalized in 1948 by Casimir and Polder \cite{cp}, from which the phenomenon took its name. By perturbation theory techniques they showed how, for the cases of the two atoms and the atom plus the wall, the behavior of the interaction energy with quantum fields, now denoted by $W^{\text{QFT}}_L$, is
\begin{equation}\label{intro_decayQFT}
W_L^{\text{QFT}} \simeq 
\begin{dcases}
-\frac{D_1(\alpha)}{L^7}, \qquad &\text{for two hydrogen atoms,}\\ 
-\frac{D_2(\alpha)}{L^4}, \qquad &\text{for a hydrogen atom and a wall},
\end{dcases}
\end{equation}
for suitable $D_1(\alpha), D_2(\alpha) >0$ in the chosen units system and a distance $L>1$ large enough.
Despite being a remarkable result of quantum field theory, the theoretical work of Casimir and Polder is not mathematically rigorous, mainly because they calculated only the first terms of the perturbative expansions of the interaction energy. Aim of the present work is to give a rigorous mathematical proof, with precise estimates, of the calculation of the interaction energy for the case of the atom plus the wall. 

The quantum nature of the van der Waals forces was first studied by London \cite{london}. The first mathematical rigorous result is due to Lieb and Thirring in \cite{LT}, where they derived an upper bound for the interaction energy between molecules, obtaining the universal $L^{-6}$ decay. The analysis was completed in \cite{anasigal} for the case of several atoms, where the correct leading order expression was derived. The literature about van der Waals interaction is extensive and includes results about further order expansions \cite{axilrod} and about interactions between various types of interfaces \cite{sernelius}. 

Casimir and Polder studied the retardation in the interaction in a non relativistic quantum electrodynamics description via a fourth (second) order expansion for the energy for two atoms (atom plus wall). At the best of our knowledge, there are many results in the physics literature continuing the line of research of Casimir and Polder (see \cite{milton} for an extensive bibliographic collection and \cite{buhmann, dalvit, dzya, salam} for monographs about the Casimir-Polder effect and van der Waals forces), but few ones with a theoretical, mathematical rigorous approach. In \cite{spohn, miyaospohn} the two atoms case is studied by the authors, who derived the decay $L^{-7}$ using a path integral formulation, making however the strong assumption that the cumulants over the second order give smaller contributions in terms of the inverse of the distance. In \cite{miyao} one of the two authors obtained again the $L^{-7}$ decay using similar techniques and estimating the higher order cumulants too, but assuming a dipole approximation and strong binding of the electrons to the nuclei (harmonic traps approximating the Coulomb attraction). Nevertheless, the cancellation of the van der Waals term of order $L^{-6}$ is not obtained by the contribution of the radiation, which is a fundamental mechanism to explain the retardation effects, as it is clear from \cite{cp}. The cancellation of the van der Waals term is recovered in Koppen's PhD thesis \cite{koppen}: the author considers a quantum electrodynamics model introducing an infrared cutoff in the Hamiltonian and studying the fourth order perturbative expansion of the energy in dipole approximation. To take  the infrared limit is, however, known to be a very difficult problem and the result is affected by the same problem of considering a truncated perturbative expansion.

Other rigorous results concern only the Casimir effect \cite{casimir} where the interaction with the matter is encoded in the boundary conditions and the radiation, described via a scalar field, is influenced only by the geometry of the classical interfaces: the vacuum energy is calculated in \cite{hasler, buenzli, fermicacciapuoti, fermi1, fermi2, fermi3}.

In \cite{cornu} the authors apply the same techniques as \cite{spohn} to the case of the atom and the wall reobtaining the behaviors \eqref{intro:decayQM} and \eqref{intro_decayQFT}, but still lacking full mathematical rigor. 

The rigorous proof of the Casimir-Polder effect for the general setting is, thus, still an open problem.

In this paper we study the Casimir-Polder effect for the case of the atom interacting with the wall. In \cite{halfspace} the van der Waals interaction energy for the electrostatic setting is rigorously computed and is coherent with the decay \eqref{intro:decayQM}:
\begin{equation}\label{intro:halfspace_qm_decay}
W^{\text{qm}}_L = -\frac{\alpha^2}{L^3} + O\Big( \frac{\alpha^2}{L^5}\Big),
\end{equation}
where $L$ is the distance between the atom's nucleus and the wall and $\alpha$ is the fine structure constant. It is a common strategy in quantum field theory to consider this as a small parameter and study expansions of the physical quantities w.r.t. $\alpha$, see \cite{pizzo, semjontranslation, semjon}, and we are going to follow the same approach providing expansions of the interaction energy in terms of $\alpha$.

In order to prove the appearance of the retardation effects and relative faster decay of the interaction after a suitable distance for the quantum fields, we consider the Pauli-Fierz model. The Pauli-Fierz model has been widely used to solve problems in non relativistic quantum electrodynamics \cite{CFO2,CFO,CFO3}. We make the following assumptions:
\begin{description}
\item[$(A1)$] the dipole approximation; 
\item[$(A2)$] reduction of the action of the Hamiltonian to the interaction with $0$ and $1$ photons between matter and field.
\end{description}  
Our approach relies on the use of precise estimates for the ground state energies of the interaction system and free systems inspired by the perturbation theory, like in \cite{semjon}, and on the calculation of line integrals on the complex plane inspired by \cite{cp}. This last step allows us to obtain the important cancellation of the van der Waals term in \eqref{intro:halfspace_qm_decay} generated by the Coulomb contribution and to derive the new leading term, as stated in the main result in Theorem \ref{thm:main}:
\begin{equation}\label{intro:result}
W^{\text{QFT}}_L \simeq -\frac{\alpha}{L^4}\aleph_{\alpha,L},
\end{equation}
which, for short distances (less than $10$ Bohr radii), gives again the decay in \eqref{intro:halfspace_qm_decay} because $\aleph_{\alpha,L} \simeq \alpha L$, while for large distances (bigger than $100$ Bohr radii) gives the $L^{-4}$ behavior predicted in \eqref{intro_decayQFT} because $\aleph_{\alpha,L}\simeq \text{const.}$, and in the intermediate region expresses the transition between the two values. The techniques used let us enlighten how the retardation effects are originated from the exchanges of one photon with the matter and the interaction with the vacuum flactuations (see the calculations in Subsection \ref{subsec:norms_evaluation}).

At the best of our knowledge, our result is the first one where the Casimir-Polder effect for the model of the atom plus the wall is proven with rigorous estimates and without recurring to infrared cut-off or to perturbative expansions. The result is, nevertheless, unsatisfactory in some aspects: one would like to drop assumptions $(A1)$ and $(A2)$ and obtain the result for the full model. Furthermore, as explained in Section \ref{sec:discussion_result}, the result gives the decay behavior of the interaction energy discussed above, but the error produced in the calculations is smaller than the leading term only up to approximately $82.5$ Bohr radii. After that distance the expression \eqref{intro:result} ceases to be the leading term because, for technical difficulties, parts of the error term are uniform on the distance. In a future paper we would like to give the result for the full, non approximated model with an error suitably dependent on the distance. 

The structure of the paper is the following:
\begin{itemize}
\item in Section \ref{sec:free} we introduce the Pauli-Fierz model, a quantum electrodynamics model describing the joint system of the hydrogen atom interacting with the radiation and we recall a result from \cite{semjon}, adapted to our approximated model, for the estimate of the ground state energy of this free system;
\item in Section \ref{sec:interaction_model} we introduce a modified version of the Pauli-Fierz model to describe the interaction between the hydrogen atom and radiation with the wall, whose construction is justified in Appendix 3, and we state our main result in Theorem \ref{thm:main}. The proof is given in the following parts: in Subsections \ref{subsec:upper} and \ref{subsec:lower} we prove upper and lower bounds, respectively, for the ground state energy of the interaction system. Then, in Subsection \ref{subsec:norms_evaluation}, we calculate the difference between the ground state energies of the interaction and free systems with the technical line integral calculations postponed in Appendix 2. 
\item In Section \ref{sec:discussion_result} we discuss the relation between the term in \eqref{intro:result} and the error terms to identify the leading term in the different regimes of distance. 
\item In Appendix 1 we collected some useful estimates on the one-photon vector $\Phi_{\#}^y$ useful through all the paper.
\end{itemize}

We make a comment here about the notation that is going to be used in the paper: $C >0$ is going to denote a positive constant which is independent of the parameters of interest $\alpha$ and $L$ and which can vary from line to line. The notation $O(\cdot)$ has to be intended in the usual sense, but we remark that we did not take track of the dependence on the ultra-violet cut-off $\Lambda$, meaning that we assume $\Lambda$ to be fixed, independent of $\alpha$ and $L$ and we are not interested in studying the ultra-violet problem. For the quadratic forms of operators we use sometimes the notation $\langle H\rangle_{\Psi} := \langle \Psi\,|\, H\,|\, \Psi \rangle$. Furthermore, for a vector $v \in \mathscr{H}$ in a Hilbert space $\mathscr{H}$, and two operators $A, B$ acting on $\mathscr{H}$, where $B$ admits an inverse, we use the following fractional notation to be interpreted as the order of operations below:
\begin{equation*}
\frac{Av}{B} = \frac{A}{B} v := B^{-1} A v.
\end{equation*}

\textbf{Acknowledgement}
This research was funded by the Deutsche Forschungsgemeinschaft (DFG, German Research Foundation) Project-ID 258734477 - SFB 1173.
I thank D. Hundertmark and S. Vugalter for the intensive discussions about the topic and important suggestions about the scaling involved in the problem during my permanence in Karlsruhe Institute of Technology. I thank I. Anapolitanos for the discussions about the van der Waals forces involved in the model studied. I further thank M. Correggi, M. Falconi and L. Morin for suggestions about the presentation of the result, and the anonymous referee, whose accurate comments really helped to improve the presentation and the rigor of the paper.

\section{Free hydrogen atom with radiation: Pauli-Fierz model}\label{sec:free}

We consider a non-relativistic, quantum, spinless electron in a hydrogen atom model, therefore interacting via an electrostatic Coulomb potential with a fixed nucleus. We study the joint system of the electron and a quantum electromagnetic field with their mutual interaction.

We fix the nucleus of the atom in position $0 \in \mathbb{R}^3$ and define the position variable of the electron to be $x = (x_1, x_2, x_3) \in \mathbb{R}^3$, so that the Hilbert space associated to the hydrogen atom model is $L^2(\mathbb{R}^3; \mathrm{d} x)$. The radiation is described in a Fock space representation 
\begin{equation}
\Gamma_s(\mathfrak{h}) = \bigoplus_{n=0}^{\infty} \mathfrak{h}^{\otimes_s n},
\end{equation}
where the $n-$th sector is associated to $n$ photons, and the one photon space is  
\begin{equation}
\mathfrak{h} := L^2(\mathbb{R}^3;\CC^2;\diff k),
\end{equation}
of square integrable functions with two components in the complex numbers associated to the two perpendicular polarization directions of the electromagnetic field. 
In our notation, we are going to denote by superscripts the components of the vectors in the sectors of the Fock space:
\begin{equation}
\Psi \in \Gamma_s(\mathfrak{h}), \qquad \Psi = (\Psi^{(0)}, \Psi^{(1)}, \Psi^{(2)}, \ldots),
\end{equation}
and denote by $\Omega := (1, 0, 0, \ldots)$ the vacuum vector.

The Hilbert space for the full system is 
\begin{equation}
\mathscr{H} = L^2(\mathbb{R}^3;dx) \otimes \Gamma_s(\mathfrak{h}).
\end{equation}
We can define operator-valued distribution for the creation and annihilation operators $$\{a_{\gamma}^{\dagger}(k), a_{\gamma}(k)\}_{\gamma = 1,2},$$ which create and destroy a photon, respectively, with frequency $k \in \mathbb{R}^3$ for each direction of polarization $\gamma$ and have the following canonical commutation relations, for $\beta, \gamma \in \{1,2\}$ and $k,h \in \mathbb{R}^3$,
\begin{equation}
[a_{\beta}(k), a_{\gamma}(h)] = 0 = [a_{\beta}^{\dagger}(k), a_{\gamma}^{\dagger}(h)],\qquad [a_{\beta}(k), a^{\dagger}_{\gamma}(h)] = \delta_{\beta,\gamma} \delta(k-h).
\end{equation}
The associated field operators are then, for any $\lambda \in \mathfrak{h}$,
\begin{equation}
a(\lambda) = \sum_{\gamma = 1,2} \int \diff k \, \overline{\lambda_{\gamma}(k)} a_{\gamma}(k), \qquad a^{\dagger}(\lambda) = \sum_{\gamma = 1,2} \int \diff k \, \lambda_{\gamma}(k) a_{\gamma}^{\dagger}(k).
\end{equation}
If the wall is at infinite distance, it does not affect the system composed by the hydrogen atom and the radiation. The dynamics is generated by the so-called \textit{Pauli-Fierz Hamiltonian}, that we denote by $H_{\infty}^{\text{PF}}$ and is formally defined by the following sum
\begin{equation}
H_{\infty}^{\text{PF}} = (P \otimes \mathbb{1} - \alpha^{1/2} A_{\infty}(x))^2 + \mathbb{1} \otimes H_f -\frac{\alpha}{|x|} \otimes \mathbb{1}, 
\end{equation}
where $\alpha$ plays the role both of the square of the charge and of the coupling between matter and field.
Here $P = i \nabla_x$ is the momentum operator for the electron and 
$$H_f = \mathrm{d} \Gamma (\omega) = \sum_{\gamma = 1,2}\int \diff k \, \omega(k) \,a_{\gamma}^{\dagger}(k) a_{\gamma}(k),  $$
is the free energy operator for the field with the usual dispersion relation for the massless photons
\begin{equation}
\omega(k) = |k|.
\end{equation}
The vector field potential  $A_{\infty} (x)$ describes the interaction between electron and field. It can be expressed as the sum
\begin{equation}
A_{\infty}(x) =  A_{\infty}^+(x) +  A_{\infty}^-(x),
\end{equation}
where
\begin{equation}
A_{\infty}^+(x) = a^{\dagger}(\lambda_{\infty}(x)), \qquad  A_{\infty}^-(x) = a(\lambda_{\infty}(x)),
\end{equation}
and $A_{\infty}^+(x)$ and $A_{\infty}^-(x)$ create and annihilate a photon with state $\lambda_{\infty}(x)$, respectively, from the interaction with an electron with position variable $x$. The expression of the form factor $\lambda_{\infty} (x)= (\lambda_{\infty,\gamma}(x))_{\gamma =1,2}, $ with $\lambda_{\infty, \gamma}\in L^{\infty}(\mathbb{R}^3;\mathfrak{h}^3)$, is given by 
\begin{equation}
\lambda_{\infty,\gamma}(x) = \frac{\chi_{\Lambda}(k)}{2\pi|k|^{1/2}} \mathbf{e}_{\gamma}(k) e^{i  kx}, \qquad \gamma =1,2,
\end{equation}
with $\chi_{\Lambda}$ being defined, for a fixed, finite $\Lambda\geq 1$, as
\begin{equation}
\chi_{\Lambda}(k) = \chi\Big( \frac{|k|}{\Lambda}\Big),   \qquad \chi(r) =  \begin{cases}
1, &\text{if} \quad  r < 1/2,\\
0, &\text{if} \quad  r > 1, \end{cases} \qquad \chi \in [0, 1],
\end{equation}
where  $\chi \in C_0^{\infty}(\mathbb{R}_+)$. In this way $\chi_{\Lambda}$ is a cut-off function for frequencies of the photons over $|k| < \Lambda$. The $(\mathbf{e}_{\gamma})_{\gamma=1,2}$ are the two polarization vectors which form with $\hat{k} = \frac{k}{|k|}$ an orthonormal basis for $\mathbb{R}^3$. The vector field can be rewritten in a formal but useful way by means of the operator-valued distributions
\begin{equation}
A_{\infty}(x) = \sum_{\gamma=1,2} \int_{\mathbb{R}^3} \mathrm{d}k \;  \frac{\chi_{\Lambda}(k)}{2\pi |k|^{1/2}} \mathbf{e}_{\gamma}(k) (a_{\gamma}(k) e^{i kx} + a^{\dagger}_{\gamma}(k) e^{-ikx}).
\end{equation} 

Since $|\alpha |\leq 1$ and $\lambda_{\infty,\gamma}, \omega^{-1/2} \lambda_{\infty, \gamma} \in L^{\infty}(\mathbb{R}^3;\mathfrak{h}^3), \gamma =1,2$, by \cite[Theorem 13.3]{spohnbook} and Kato-Rellich Theorem, the Pauli-Fierz Hamiltonian is self-adjoint on the domain $\mathscr{D}(H_{\infty}) = H^2(\mathbb{R}^3) \otimes \mathscr{D}(\mathrm{d}\Gamma(\omega))$ (for more general conditions see \cite{hiroshima}).

Assuming to work in Coulomb gauge, expressed by the condition $\nabla_x \cdot A(x)= 0$, the Pauli-Fierz Hamiltonian can be rewritten, calculating the square, in the following way: 
\begin{equation}
H_{\infty} = h_{\alpha} + H_f -  2\alpha^{1/2}\mathrm{Re} (P A_{\infty}(x)) + \alpha A^2_{\infty}(x),
\end{equation}
where, from now on, we drop the tensor products with the identity in order to ease the notation. The $h_{\alpha}$ is the hydrogen atom Hamiltonian 
\begin{equation}
h_{\alpha} = -\Delta_x -\frac{\alpha}{|x|}, \qquad u_{\alpha}(x) = \frac{\alpha^{3/2}}{\sqrt{8\pi}} e^{-\alpha\frac{|x|}{2}}, \qquad e_{\alpha} = -\frac{\alpha^2}{4},
\end{equation}
with $u_{\alpha}$ and $e_{\alpha}$ being the ground state and ground state energy, respectively. When needed, we are going to use as well the notation $h_1 := h_{\alpha =1}, e_1:= e_{\alpha = 1}, u_{1} := u_{\alpha = 1}$.

As we anticipated, we are going to work with the dipole approximated Hamiltonian whose action is restricted to $0$ and $1$ photons. In order to do that we introduce the projector on the first sectors of the Fock space
\begin{align*}
\Pi: \Gamma_s(\mathfrak{h})  &\longrightarrow \mathbb{C} \oplus \mathfrak{h}\\
\Psi  &\longmapsto (\Psi^{(0)}, \Psi^{(1)}),
\end{align*}
whose action on pure tensors is 
\begin{equation}
\Pi (f \otimes \Psi) := f\otimes \Pi \Psi = f \otimes (\Psi^{(0)}, \Psi^{(1)}), \qquad f \in L^2(\mathbb{R}^3;dx), \Psi \in \Gamma_s(\mathfrak{h}).
\end{equation}
We apply the substitution below on the new Hamiltonian $\Pi H_{\infty} \Pi^{\dagger}$, called \textit{dipole approximation},
\begin{equation}\label{eq:dipole_approx}
A_{\infty}^{\pm}(x) \quad \longmapsto \quad  A_{\infty}^{\pm}(0) =: A_{\infty}^{\pm} 
\end{equation}
which makes the argument of the creation and annihilation operators to be 
\begin{equation}\label{eq:lambda_0_def}
\lambda_{\infty} = (\lambda_{\infty,\gamma})_{\gamma =1,2}, \qquad \lambda_{\infty,\gamma} := \lambda_{\infty,\gamma}(0) =  \frac{\chi_{\Lambda}(k)}{2\pi|k|^{1/2}} \mathbf{e}_{\gamma}(k),
\end{equation} 
obtaining the new approximated, free Hamiltonian
\begin{equation}\label{eq:free_approx_PF}
H_{\infty} :=  h_{\alpha} + H_f -2 \alpha^{1/2} \mathrm{Re}PA_{\infty} + \alpha \|\lambda_{\infty}\|^2 + 2 \alpha A_{\infty}^{+} A_{\infty}^-,  
\end{equation}
acting on the space
\begin{equation}
\mathscr{H}_{\infty} := \Pi \mathscr{H}  = L^2(\mathbb{R}^3;dx) \otimes (\mathbb{C} \oplus \mathfrak{h}). 
\end{equation}
We observe that the third term in \eqref{eq:free_approx_PF} is the only one which changes the number of photons.
Let us further denote by 
\begin{equation}
E_{\infty} := \inf \sigma (H_{\infty}),
\end{equation}
the ground state energy of the approximated, free Hamiltonian. We are now ready to state an adaptation of the result \cite[Theorem 2.1]{semjon} in our setting with at most one photon. 
Let us introduce the following scalar products on $\mathscr{H}_{\infty}$,
\begin{equation}\label{eq:scal_prod_def}
\langle \cdot \,|\, \cdot\rangle_{\#} :=\langle \cdot\,|\, (h_{\alpha}- e_{\alpha} + H_f)\,|\, \cdot \rangle, \qquad 
\langle \cdot\,|\, \cdot \rangle_{*}:= \langle \cdot \,|\, H_f \,|\,\cdot\rangle, 
\end{equation}
and the vectors 
 \begin{align*}
\Phi_{\#}^{\infty} :=  2 \alpha^{1/2} (h_{\alpha}- e_{\alpha} +H_f)^{-1} P u_{\alpha} \otimes  A_{\infty}^+ \Omega, \qquad 
\Phi_*^{\infty} := 2 \alpha^{1/2} P u_{\alpha} \otimes H_f^{-1} A_{\infty}^+ \Omega,
\end{align*}
where the second one is not a vector belonging to the Hilbert space, but it is going to appear only in expressions which make sense. We also define the following vector 
\begin{align*}
\Phi_1^* = H_f^{-1} P_f A_{\infty}^- H_f^{-1} A_{\infty}^+A_{\infty}^+ \Omega, 
\end{align*}
where $P_f := d \Gamma (k)$ is the momentum operator for the field. 

\begin{theorem}\label{thm:freebound}
The following estimate holds for the energy of the free Hamiltonian,
\begin{equation}
E_{\infty} = e_{\alpha} + \alpha \|\lambda_{\infty}\|^2  - \|\Phi_{\#}^{\infty}\|^2_{\#}  - 4\alpha^3  \|\Phi_1^*\|^2_{*} +  O(\alpha^4 \log(\alpha^{-1})).
\end{equation}
\end{theorem}

\begin{proof}
From \cite[Theorem 5.1]{semjon} we get the upper bound, adapted for the Hamiltonian with $0$ and $1$ photons,
\begin{equation}
E_{\infty} \leq e_{\alpha} + \alpha \|\lambda_{\infty}\|^2  - \|\Phi_{\#}^{\infty}\|^2_{\#}  - 4\alpha^3  \|\Phi_1^*\|^2_{*} +  O(\alpha^4 \log(\alpha^{-1})),
\end{equation}
by choosing a suitable trial function. The term $\alpha \|\lambda_{\infty}\|^2$ appears because we do not consider the normal ordered Hamiltonian like in the cited paper.
We observe that instead of having an error of order $O(\alpha^4)$ we get a $O(\alpha^4 \log(\alpha^{-1}))$ term because of the appearance of the additional term $\alpha \|\lambda_{\infty}\|^2 \|\Phi_{\#}^{\infty}\|^2$ in the calculations compared to the original version, which is treated in a similar way as its analogous in the interaction model (see formula \eqref{eq:quad_Phiy_ort}).

By \cite[Theorem 5.2]{semjon} we obtain the lower bound
\begin{equation}
E_{\infty} \geq e_{\alpha} + \alpha \|\lambda_{\infty}\|^2  - \|\Phi_{*}^{\infty}\|^2_{*}  - 4\alpha^3  \|\Phi_1^*\|^2_{*} +  O(\alpha^4 \log(\alpha^{-1})).
\end{equation}
The substitution of $\|\Phi_*^{\infty}\|_*^2$ with $\|\Phi_{\#}^{\infty}\|^2_{\#}$ produces the error term, thanks to \cite[Lemma C.5]{semjon},
\begin{equation}
\|\Phi_*^{\infty}\|_*^2 - \|\Phi_{\#}^{\infty}\|^2_{\#} = O(\alpha^{5}\log(\alpha^{-1})),
\end{equation}
which is reabsorbed in the error term $O(\alpha^4 \log (\alpha^{-1})).$
\end{proof}

\section{Interaction model: atom and wall}\label{sec:interaction_model}

We are now ready to define the interaction Hamiltonian, which shares, except for the presence of the Coulomb potential with the wall, the same structure with the free Hamiltonian, but in the vector potential it is clear how the presence of the wall influences the energy.
Without loss of generality we can consider the wall to be parallel to the plane $\Sigma_0 =\{(0,x_2,x_3)\,|\, x_2, x_3 \in \mathbb{R}\}$, translated in the $x_1$ direction by a distance $y > 0$ in the positive semi-line, so that the conductor wall is described by $\Sigma_y =\{(y, x_2, x_3) \,|\, x_2, x_3 \in \mathbb{R}\}$. 

The space for the particle is set to be 
\begin{equation}
L^2(\mathbb{R}_y^3;dx) \quad  \text{where}, \quad 
\mathbb{R}_y^3 = \{x = (x_1, x_2, x_3) \in \mathbb{R}^3\,|\, x_1 < y\}.
\end{equation}
We express the distance as a multiple of the Bohr radius, given by the inverse of the fine structure constant $\alpha$ to make it homogeneous with the physical quantities we are going to introduce in the following, so that 
\begin{equation}\label{eq:y_rel_alpha}
y = L \alpha^{-1}, \qquad L >1.
\end{equation}
By an abuse of notation, we denote by $y$ both the length \eqref{eq:y_rel_alpha} and the vector
\begin{equation}
y = (L\alpha^{-1},0,0),
\end{equation}
the choice being clear from the context.
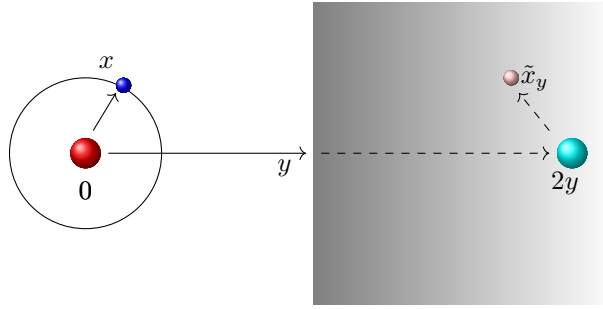
\begin{figure}
\caption{Interaction described by the image charge method.}
\begin{center}
\begin{tikzpicture}
\begin{scope}[yscale=1,xscale=-1]
\shade[left color=gray, right color=white] (-5,-2) rectangle (-1,2);
\shade[shading=ball, ball color=red] (2,0) circle (.2); \node at (2,-0.5){$0$}; 
\draw (2,0) circle (1); 
\shade[shading=ball, ball color = blue] (1.5,0.9) circle (.1); \node[above, left] at (1.5,1.2) {$x$};
\draw[->] (1.7,0) -- (-0.9,0);\node[below, right] at (-0.4,-.2) {$y$};
\draw[->] (1.9,0.3) -- (1.6,0.8); 
\draw[dashed, ->] (-1.1,0) -- (-4.1,0);\node[below, right] at (-4,-.4) {$2y$};
\draw[dashed, ->] (-4.1,0.3) -- (-3.7,0.8); 
\shade[shading=ball, ball color= cyan] (-4.4,0) circle (.2); \node at (2,-0.5){0};
\shade[shading=ball, ball color = pink] (-3.6,1) circle (.1); \node[right] at (-3.6, 1){$\tilde{x}_y $};
\end{scope}
\end{tikzpicture}
\end{center}
\end{figure}

The Coulomb interaction with the wall is equivalent, thanks to the well known image charge method, to the interaction with a mirror atom with inverted charges:
\begin{equation*}
V_y(x) = \frac{1}{2} \left( -\frac{1}{2|y|}+\frac{1}{|\tilde{x}_y|} +\frac{1}{|x-2y|}- \frac{1}{|\tilde{x}_y - x|}\right), \quad \tilde{x}_y := (2y - x_1, x_2, x_3).
\end{equation*}
By \cite[Lemma 2.1]{halfspace} we know that $V_y\leq 0$. 
For future purposes, we make the following split of the potential
\begin{equation}
V_y^{>} :=- \frac{1}{2|\tilde{x}_y - x|}, \qquad V_y^{<} := V_y-V_y^{>},
\end{equation}
and observe that, in $\mathbb{R}^{3}_y$, the following bounds hold: there exists a $C>0$ such that 
\begin{align}
&|V_y^{<}(x)| \leq \frac{C}{y},\qquad &\text{for any } x \in \mathbb{R}^3_y, \label{eq:V<_prop}\\
\int_{\mathbb{R}^3_y} \diff x\, &(V^{>}_y(x))^2 |u(x)|^2 \leq C \int_{\mathbb{R}^3_y} \diff x\, |P u(x)|^2, \qquad &\text{for any } u \in H^1_0(\mathbb{R}^3_y),\label{eq:V>_prop}
\end{align} 
the second one being a Hardy-type inequality proven in \cite[Lemma 3.1]{halfspace}.

The electromagnetic field is described by the Fock space with photons with positive frequencies in the direction normal to the wall:
\begin{equation}
\Gamma_s(\mathfrak{h}_+)= \bigoplus_{n=0}^{\infty} \mathfrak{h}_+^{\otimes_s n}, \qquad \mathfrak{h}_+ := L^2(\mathbb{R}^+ \times \mathbb{R}^2; \mathbb{C}^2; \mathrm{d}k),
\end{equation}
where the two polarization directions of the photons are taken into account. The full Hilbert space is, in this case,
\begin{equation}
\mathscr{H}_y = L^2(\mathbb{R}^3_y) \otimes \Gamma_s(\mathfrak{h}_+),
\end{equation}
and the Hamiltonian generating the dynamics is formally given by the expression
\begin{equation}
H_y^{\text{PF}} := h_{\alpha} + H_f^+ - 2 \alpha^{1/2} \mathrm{Re} (P A_{y}(x)) + \alpha A^2_{y}(x) + \alpha V_y(x),
\end{equation} 
where the free field energy is 
\begin{equation}
H_f^+ := \sum_{\gamma =1,2} \int_{\mathbb{R}^+ \times \mathbb{R}^2} \diff k\, \omega(k) a^{\dagger}_{\gamma}(k) a_{\gamma}(k).
\end{equation}
Here again we can split $A_y$ in creation and annihilation parts
\begin{equation}
A_{y}(x) =  A_{y}^+(x) +  A_{y}^-(x),
\end{equation}
where 
\begin{equation}
 A_{y}^+(x)= a^{\dagger}(\lambda_{y}(x)), \qquad  A_{y}^-(x) = a(\lambda_{y}(x)),
\end{equation}
the form factor $\lambda_y(x) = (\lambda_{y,\gamma}(x))_{\gamma =1,2}$, with $\lambda_{y,\gamma}\in L^{\infty}(\mathbb{R}^3_y;\mathfrak{h}^3_+)$, this time being
\begin{equation}\label{eq:formfactorinter}
\lambda_{y,\gamma}(x) =\frac{\chi_{\Lambda}(k)}{2\pi |k|^{1/2}} e^{i(k_2 x_2 + k_3 x_3)} \left( \begin{array}{c}
\mathbf{e}^{(1)}_{\gamma}(k) 2\cos(k_1 (x_1 - y)) \\
\mathbf{e}^{(2)}_{\gamma}(k)2i \sin (k_1 (x_1 - y)) \\
\mathbf{e}^{(3)}_{\gamma}(k) 2i  \sin (k_1 (x_1-y))\end{array}
\right).
\end{equation}
Here by $\mathbf{e}_{\gamma}^{(j)}$ is the $j-$th component of the $\gamma-$th polarization vector.

In Appendix 3 we give a justification of the definition of this Hamiltonian as the right one to describe the model of the atom interacting with the wall.
Theorem 5.7 in \cite{matte} ensures the self-adjointness of the Hamiltonian, provided that the following conditions are satisfied: following the notation of the mentioned paper, we choose $\mathcal{M} = \mathbb{R}_+ \times \mathbb{R}^2$; $H = H_y^{\text{PF}}$; $\omega(k) = |k|$; $\lambda = \lambda_y$; $V = -\frac{\alpha}{|x|} + \alpha V_y(x)$. In particular, recalling that $\tilde{x}_{y} = (2y - x_1,  x_2,  x_3)$, 
\begin{align*}
\nabla \cdot \lambda_{y,\gamma} &= \frac{\chi_{\Lambda}(k)}{\pi |k|^{1/2}}\{\mathbf{e}^{(1)}_{\gamma}(k) \der_{x_1}( e^{i (k_2 x_2 + k_3 x_3)} \cos(k_1(x_1 - y)) )\\
 &\qquad \quad \;\;+
\mathbf{e}^{(2)}_{\gamma}(k) \der_{x_2}( e^{i (k_2 x_2 + k_3 x_3)}i \sin(k_1(x_1 - y)) )  \\
&\qquad \quad \;\;+\mathbf{e}^{(3)}_{\gamma}(k)\der_{x_3}( e^{i (k_2 x_2 + k_3 x_3)} i\sin(k_1(x_1 - y)) )\}   \\
&= - \frac{\chi_{\Lambda}(k)}{\pi |k|^{1/2}}k \cdot \mathbf{e}_{\gamma} (k) \, e^{i (k_2 x_2 + k_3 x_3)} \sin(k_1(x_1 - y))  = 0, \qquad \gamma = 1,2.
\end{align*}
Therefore \cite[Theorem 5.7]{matte} can be applied and $H_y^{\text{PF}}$ is self-adjoint on 
\begin{equation}
\mathscr{D}(H^{\text{PF}}_y) = \mathscr{D}((-\Delta^D) \otimes \mathbb{1}) \cap \mathscr{D}(\mathbb{1} \otimes \mathrm{d}\Gamma (|k|)),
\end{equation}
where $-\Delta^{D}$ is the Dirichlet Laplacian.
As for the free model, we reduce the action of this Hamiltonian to the $0$ and $1$ photons space. By an abuse of notation, we denote again by $\Pi$ the projector over the $0-$th and $1-$st Fock sectors of $\Gamma_s(\mathfrak{h}_+)$ and apply an analogous dipole approximation as \eqref{eq:dipole_approx} for the $A^{\pm}_y$ in $\Pi H_y^{\text{PF}}\Pi^{\dagger}$ to obtain
\begin{equation}
H_y := h_{\alpha} + H_f^+ + \alpha V_{y} -2 \alpha^{1/2} \mathrm{Re}PA_{y}(x) + \alpha \|\lambda_{y}\|^2 + 2 \alpha A_{y}^{+} A_{y}^-,
\end{equation}
where now the argument of the creation and annihilation operators $A_y^{\pm}$ has the form
\begin{equation}\label{eq:lambda_y_0_def}
\lambda_y = (\lambda_{y,\gamma})_{\gamma=1,2}, \qquad \lambda_{y,\gamma} := \lambda_{y,\gamma}(0) =  \frac{\chi_{\Lambda}(k)}{2\pi |k|^{1/2}} \left( \begin{array}{c}
\mathbf{e}^{(1)}_{\gamma}(k)2 \cos(k_1 y) \\
 -\mathbf{e}^{(2)}_{\gamma}(k) 2i\sin (k_1 y) \\
-\mathbf{e}^{(3)}_{\gamma}(k)  2i\sin (k_1 y)\end{array}
\right) .
\end{equation}

Introducing the ground state energy 
\begin{equation}
E_y := \inf \sigma (H_y),
\end{equation}
we estimate it in the next theorem.
\begin{theorem}\label{thm:estimate_interaction}
For any $L>1$, we have 
\begin{equation}
E_y = e_{\alpha} - \frac{\alpha^2}{L^3} + \alpha \|\lambda_y\|^2 - \|\Phi^{\#}_y\|^2_{\#} + O\Big(\frac{\alpha^2}{L^5} \Big) + O(\alpha^4 \log(\alpha^{-1}))+ O\big( \alpha^2 L e^{-L/2}\big). 
\end{equation}
\end{theorem}

The proof consists in giving upper and lower bounds, which is the content of Subsections \ref{subsec:upper} and \ref{subsec:lower}, respectively.

We use the estimates for the free energy $E_{\infty}$, the estimates for the energy of the interaction system $E_y$ from Theorems \ref{thm:freebound} and \ref{thm:estimate_interaction}, respectively, and the important estimates from Proposition \ref{propos:E_norm_estimates} in Subsection \ref{subsec:norms_evaluation}, to prove the main theorem of the paper.
\begin{theorem}\label{thm:main}
For any $L>1$, we have
\begin{align}
W_y^{QFT} &= E_y -E_{\infty} \nonumber\\
&= -\aleph_{\alpha,L} \frac{\alpha}{L^4}+  O(\alpha^4 \log(\alpha^{-1}))+ O\big( \alpha^2 L e^{-L/2}\big) + O \Big( \frac{\alpha^2}{L^5}\Big) + O\Big( \frac{\alpha^3}{L^2}\log(\alpha^{-1})\Big), 
\end{align}
where 
\begin{equation}
\aleph_{\alpha,L} :=  \frac{1}{6 \pi } \left\langle\alpha L\arctan\left(\frac{1}{\alpha L (h_1 - e_1)}\right) \right\rangle_{x u_1}, 
\end{equation}
and we have the following behaviors depending on the chosen regime for the distance:
\begin{equation}\label{eq:aleph_def}
  \aleph_{\alpha,L} \simeq \begin{dcases}
        \alpha L, \quad &\text{if} \quad 1< L \leq \frac{16}{3}, \\
        \frac{16}{3}\alpha^{\eta}, \quad &\text{if} \quad L = \frac{16}{3} \alpha^{-1+\eta}, \; \eta \in (0,1),\\
       \frac{1}{6\pi}\|(h_1-e_1)^{-1/2}x u_1\|^2, \quad &\text{if} \quad L \geq \frac{16}{3} \alpha^{-1}.
        \end{dcases}
 \end{equation}
\end{theorem}

The interpretation of the result is going to be studied in Section \ref{sec:discussion_result}.

\begin{remark}
In the error terms in Theorem \ref{thm:main} it is not stated explicitly the dependence on the ultraviolet cut-off $\Lambda$. Some of the bounds depend, indeed, on the choice of $\Lambda$ that anyway we keep fixed and independent on $\alpha$ and $L$, and therefore they should be improved if one wanted to study the removal of the cut-off. Therefore, the estimates hold considering $\alpha$ as a small parameter compared to the other quantities and $\Lambda$ as a fixed constant value. Since we are not interested in studying the ultra-violet problem and since the key quantities come from the interactions with photons with low momenta, we may set for simplicity $\Lambda =1$ to keep it independent on the numerical value of $\alpha$.
\end{remark}

\begin{remark}
The cancellation of the van der Waals term and the leading term in the model are obtained considering only interactions with zero and one photons. It is therefore reasonable that the interactions with a higher number of photons and without considering the dipole approximation would contribute only by error terms. The technical difficulties in doing so consist in a heavier computational cost (the number of terms to bound is really higher), in dealing with the position-dependence on the field operators and on the fact these bounds require the derivation of number estimates for the states of minimal energy of the interaction model. 
This is what we aim to prove in a future paper.
\end{remark}

\subsection{Upper bound}\label{subsec:upper}
In this subsection we are going to prove, in the theorem below, an upper bound for $E_y$ providing in this way the first step of the proof for Theorem \ref{thm:estimate_interaction}.
We use the convention, for $f,g \in L^2(\mathbb{R}^3_y)$ and $\Psi, \Phi \in \mathfrak{h}_+$,
\begin{equation}
\langle f\otimes \Psi\,|\, g \otimes \Phi\rangle = \sum_{\gamma = 1,2}\int_{\mathbb{R}^3_y} dx \int_{\mathbb{R}^3_+} \diff k\, \overline{f(x)} g(x)\overline{\Psi_{\gamma}(k)} \Phi_{\gamma}(k). 
\end{equation}
\begin{theorem}\label{thm:upper_bound}
There exists a function $\varphi_y \in \mathscr{D}(H_y)$ such that, for any $L>1$,
\begin{align}
\frac{\langle \varphi_y\,|\, H_y\,|\, \varphi_y\rangle}{\langle \varphi_y\,|\,\varphi_y\rangle} &\leq e_{\alpha} - \frac{\alpha^2}{L^3} + \alpha \|\lambda_y\|^2 - \|\Phi^{\#}_y\|^2_{\#} \nonumber \\
&\quad +4 \alpha^3 \|\Phi^1_*\|^2_*+ O\Big(\frac{\alpha^2}{L^5} \Big) + O(\alpha^4 \log(\alpha^{-1})) + O(\alpha^2 L e^{-L/2}). 
\end{align}
\end{theorem}
In order to prove the theorem we construct the trial function $\varphi_y$ in the following way: we define the vector $\Phi_{\#}^y$ in an analogous way as we did for the relative free version: 
\begin{align*}
\Phi_{\#}^y := 2 \alpha^{1/2} (h_{\alpha}-e_{\alpha} + H_f^+)^{-1}P u_{\alpha} \otimes  A_{y}^+\Omega.
\end{align*}
We then introduce the trial function
\begin{align*}
\varphi_y :=  u_{\alpha}\otimes (\Omega + 2\alpha^{3/2} \tilde{\Phi}_{*}^1) + \Phi_{\#}^y,
\end{align*}
where the vector $\tilde{\Phi}_*^1 := \sqrt{2} \Phi_*^{1}|_{k \in \mathbb{R}^3_+}$ is the restriction of $\Phi_*^1$ to the positive $k_1$ frequencies.
We calculate the norm of the trial function.
\begin{lemma}\label{lem:norm_trial_function}
The trial function $\varphi_y$ has the following norm
\begin{equation}
\|\varphi_y\|^2 = 1 + O(\alpha^3\log(\alpha^{-1})).
\end{equation}
\end{lemma}
\begin{proof}
Since the vacuum vector and the last addend composing the trial function live in two different Fock sectors, since $u_{\alpha} \bot \Phi_{\#}^y$ and $u_{\alpha} \otimes \Omega$ has norm $1$, we can write
\begin{align*}
\|\varphi_y\|^2 &= 1 + 4 \alpha^3\|\tilde{\Phi}_*^1\|^2 + \|\Phi_{\#}^y\|^2, 
\end{align*}
we can conclude by applying Lemma \ref{lem:norm_vector_ort} to the last term.
\end{proof}

In the following we are going to make use of the exponential decay of the ground state of the hydrogen atom reformulated in the next lemma to exploit the dependence on the parameters for our setting. We introduce $\zeta_y \in C^{\infty}_0(\mathbb{R}^3)$, a smooth, radial characteristic function, with 
\begin{equation}
\zeta_y (x) = \begin{cases}
1,  &\text{for} \quad  |x| \leq \frac{1}{4}y,\\
0, &\text{for} \quad  |x| \geq \frac{1}{3}y, 
\end{cases}
\end{equation}
which localizes the electron in a neighborhood of the origin strictly smaller than the distance from the wall.
\begin{lemma}\label{lem:gs_hyd_decay}
There exists $C >0$ such that the following holds, for any $L>1$,
\begin{equation}
\|u_{\alpha}(1-\zeta_y)\|^2  \leq C L^2 e^{-L}.
\end{equation}
\end{lemma}
The proof is a straightforward direct calculation of the norm. Localizing in a neighborhood of zero we can consider the Taylor expansion of the potential $\alpha V_y$:
\begin{equation}\label{eq:taylor_potential}
 V_y(x) = -\frac{(x\cdot\hat{y})^2 + |x|^2}{8 y^3} +\frac{f_{\text{odd}}(x)}{8y^4} + O\Big( \frac{|x|^4}{y^5}\Big), \qquad \text{for any } x \in B_R(0),\; R>0,
\end{equation}
where $f_{\text{odd}}$ is an odd function in $x$ such that $|f_{\text{odd}}(x)|\leq C |x|^3$.
 A direct consequence is that, for the Coulomb potential of interaction with the wall we have the estimate, recalling that $y = L \alpha^{-1}$,
\begin{align}
\langle u_{\alpha}\,|\, \alpha V_y\,|\, u_{\alpha} \rangle &= \langle u_{\alpha}\,|\,\zeta_y \alpha V_y\,|\, u_{\alpha} \rangle + \langle u_{\alpha}\,|\,(1-\zeta_y) \alpha V_y\,|\, u_{\alpha} \rangle \nonumber \\
&\leq -\frac{\alpha^2}{L^3} + O\Big( \frac{\alpha^2}{L^5}\Big) +  \alpha\| (1-\zeta_y) u_{\alpha}\| \Big( \|V_y^{>} u_{\alpha} \| + \frac{C}{y} \Big) \nonumber \\
&= -\frac{\alpha^2}{L^3} + O\Big( \frac{\alpha^2}{L^5}\Big) + O\big( \alpha^2 L e^{-L/2}\big),\label{eq:potential_out_loc}
\end{align}
where for the localized part we used \eqref{eq:taylor_potential}, while for the complementary part we used Lemma \ref{lem:gs_hyd_decay}, \eqref{eq:V<_prop} and \eqref{eq:V>_prop}, the last one giving 
\begin{equation}\label{eq:V>u_prop}
\|V_y^{>} u_{\alpha}\| \leq C\|P u_{\alpha}\| \leq C \alpha.
\end{equation}

We are now ready to prove Theorem \ref{thm:upper_bound}.

\begin{proof}[of Theorem \ref{thm:upper_bound}]
Let us calculate the quadratic form of the Hamiltonian on the function $\varphi_y$
\begin{align}
\langle H_y \rangle_{\varphi_y} &= \langle H_y \rangle_{u_{\alpha} \otimes \Omega} + \langle H_y \rangle_{u_{\alpha} \otimes 2\alpha^{3/2}\tilde{\Phi}_*^1} + \langle H_y \rangle_{\Phi^y_{\#}} + 2 \mathrm{Re} \langle u_{\alpha} \otimes \Omega\,|\,H_y  u_{\alpha} \otimes 2\alpha^{3/2}\tilde{\Phi}_*^1\rangle \nonumber\\
&\quad + 2 \mathrm{Re} \langle u_{\alpha} \otimes \Omega\,|\,  H_y\Phi^y_{\#}\rangle + 2 \mathrm{Re} \langle u_{\alpha} \otimes 2 \alpha^{3/2}\tilde{\Phi}_*^1\,|\,  H_y \Phi^y_{\#}\rangle. \label{eq:quadratic_form_trial}
\end{align}
For the reader's convenience, we recall the Hamiltonian expression
\begin{equation}
H_y = h_{\alpha} + H_f^+ + \alpha V_{y} -2 \alpha^{1/2} \mathrm{Re}PA_{y}(x) + \alpha \|\lambda_{y}\|^2 + 2 \alpha A_{y}^{+} A_{y}^-.
\end{equation}

We define similar scalar products as \eqref{eq:scal_prod_def}, but in the interaction case and denote them in the same way by an abuse of notation:
\begin{equation}
\langle \cdot \,|\, \cdot\rangle_{\#} :=\langle \cdot\,|\, (h_{\alpha}- e_{\alpha} + H_f^+)\,|\, \cdot \rangle, \qquad 
\langle \cdot\,|\, \cdot \rangle_{*}:= \langle \cdot \,|\, H_f^+ \,|\,\cdot\rangle.
\end{equation}

Let us calculate each term separately. For the quadratic form in $u_{\alpha}\otimes \Omega$ we have, using \eqref{eq:potential_out_loc},
\begin{align}
\langle H_y \rangle_{u_{\alpha} \otimes \Omega} &= e_{\alpha} + \alpha\|\lambda_y  \|^2  + \langle \alpha V\rangle_{u_{\alpha}} \nonumber\\
&= e_{\alpha} + \alpha\|\lambda_y  \|^2 -\frac{\alpha^2}{L^3} + O\Big( \frac{\alpha^2}{L^5}\Big) + O\big( \alpha^2 L e^{-L/2}\big).\label{eq:quad_uOmega}
\end{align}
Let us consider the quadratic form in $u_{\alpha}\otimes 2 \alpha^{3/2} \tilde{\Phi}_*^1$:
\begin{align}
\langle H_y \rangle_{u_{\alpha}\otimes 2 \alpha^{3/2} \Phi_*^1} &= 4\alpha^3 \Big( e_{\alpha} + \alpha \|\lambda_y \|^2  + \|(H_f^+)^{1/2}\tilde{\Phi}_*^1\|^2 + 2 \alpha \|A^-_y \tilde{\Phi}_*^1\|^2\Big) \nonumber \\
&= 4 \alpha^3 \|\tilde{\Phi}^1_*\|^2_* + O(\alpha^4) = 4 \alpha^3 \|\Phi^1_*\|^2_* + O(\alpha^4),\label{eq:quad_Phi1}
\end{align}
where in the last step we used the definition of $\tilde{\Phi}_*^1$ and that $\|\tilde{\Phi}^1_*\|^2_*$ is an integral of an even function on the half plane.
For the quadratic form in $\Phi_{\#}^y$ we have
\begin{align}
\langle H_y \rangle_{\Phi_{\#}^y} &= \|\Phi_{\#}^y\|_{\#}^2 + (e_{\alpha}+\alpha\|\lambda_y\|^2) \|\Phi_{\#}^y\|^2 + \alpha \langle V_y\rangle_{\Phi_{\#}^y} + 2 \alpha \|A_y^- \Phi_{\#}^y\|^2 \nonumber\\
&\leq \|\Phi_{\#}^y\|_{\#}^2 + O(\alpha^4 \log(\alpha^{-1})), \label{eq:quad_Phiy_ort}
\end{align}
where we used \eqref{eq:norm_Phi_ort} , \eqref{eq:norm_A-Phi_ort} and we bounded from above $V_y \leq 0$.

For the cross terms
\begin{equation}\label{eq:cross_u_Phi_ort}
2 \mathrm{Re} \langle u_{\alpha} \otimes \Omega\,|\, H_y u_{\alpha} \otimes 2\alpha^{3/2}\Phi_*^1\rangle = - 4 \alpha^2 \mathrm{Re} \langle P u_{\alpha} \otimes A_y^+\Omega\,|\,u_{\alpha} \Phi_*^1 \rangle = 0, 
\end{equation}
where the other terms vanished because products between objects in two different Fock sectors and the last one remaining is zero because it is a scalar product between an odd and an even function in position variable. The same happens for 
\begin{align}
2 \mathrm{Re} \langle u_{\alpha} \otimes \Omega\,|\,  H_y\Phi^y_{\#}\rangle  &= - 8 \alpha \langle P u_{\alpha} \otimes A^+_y \Omega \,|\, h_{\alpha} - e_{\alpha}+ H_f^+\,|\,P u_{\alpha} \otimes A^+_y \Omega \rangle  \nonumber\\
&= - 2 \|\Phi_{\#}^y\|^2_{\#}. \label{eq:Phi_ort_negative}
\end{align}
For the last term it holds
\begin{equation}\label{eq:u_Phi1_Phi_ort}
2 \mathrm{Re} \langle u_{\alpha} \otimes 2 \alpha^{3/2}\tilde{\Phi}_*^1\,|\,  H_y \Phi^y_{\#}\rangle =   O(\alpha^4),
\end{equation}
where we used that 
\begin{align*}
 2 \alpha^{5/2} \mathrm{Re}\langle  u_{\alpha} \tilde{\Phi}_*^1\,|\, (\|\lambda_y\|^2 + A^+_y A^-_y) \Phi_{\#}^y \rangle &= 0,\\
 8 \alpha^2 \mathrm{Re} \langle u_{\alpha} \tilde{\Phi}_*^1\,|\, P u_{\alpha} A^+_y \Omega \rangle &= 0
\end{align*}
because $u_{\alpha}$ is orthogonal in $L^2$ to $P u_{\alpha}$ and $\Phi_{\#}^y$ and that, by \eqref{eq:taylor_potential},
\begin{equation}\label{eq:Phi1_Phiort}
4 \alpha^{5/2} | \mathrm{Re}\langle  u_{\alpha} \tilde{\Phi}_*^1\,|\, V_y \Phi_{\#}^y \rangle| \leq C \alpha^{5/2} \|\tilde{\Phi}_*^1\| \|V_y u_{\alpha}\|  \|\Phi_{\#}^y\| = O(\alpha^4),
\end{equation}
where we also used Lemma \ref{lem:norm_vector_ort} and the combination of \eqref{eq:V<_prop} and \eqref{eq:V>_prop}. 

Plugging now \eqref{eq:quad_uOmega}, \eqref{eq:quad_Phi1}, \eqref{eq:quad_Phiy_ort}, \eqref{eq:cross_u_Phi_ort}, \eqref{eq:Phi_ort_negative} and \eqref{eq:u_Phi1_Phi_ort} in \eqref{eq:quadratic_form_trial} gives the desired upper bound already. Since, thanks to Lemma \ref{lem:norm_trial_function} the contribution of the norm of the trial function is
\begin{equation}
\|\varphi_y\|^{-2} \simeq 1+ O(\alpha^3\log(\alpha^{-1})), 
\end{equation}
we see that it leaves invariant the upper bound on \eqref{eq:quadratic_form_trial} because $O(\alpha^3\log(\alpha^{-1}))$ multiplied with the leading term gives a $O(\alpha^4)$ contribution, proving the theorem.
\end{proof}

\subsection{Lower bound}\label{subsec:lower}
In this subsection we give the second and final step for the proof of Theorem \ref{thm:estimate_interaction}, giving a lower bound of $E_y$, content of the Theorem below.
\begin{theorem}\label{thm:lower_bound}
For any $L>1$,
\begin{align}
E_y &\geq e_{\alpha} - \frac{\alpha^2}{L^3} + \alpha \|\lambda_y\|^2 - \|\Phi^{\#}_y\|^2_{\#} +4 \alpha^3 \|\Phi^1_*\|^2_* \nonumber \\
&\quad + O\Big(\frac{\alpha^2}{L^5} \Big) + O(\alpha^4 \log (\alpha^{-1})) + O\big( \alpha^2 L e^{-L/2}\big). 
\end{align}
\end{theorem}

\begin{proof}
Let $\Psi_y$ denote the normalized ground state of $H_y$ so that 
\begin{equation}
E_y = \langle \Psi_y\,|\,H_y \,|\,\Psi_y\rangle.
\end{equation}
We decompose $\Psi_y$ in the following way:
\begin{equation}
\Psi_y = u_{\alpha} \otimes \Phi_y +  R_y,
\end{equation}
where
\begin{itemize}
\item $\Phi_y := \langle u_{\alpha}\,|\,\Psi_y\rangle_{L^2(\mathbb{R}^3_y)}$ and we further decompose
\begin{equation}
\Phi_y = \Phi_y^{(0)} + 2 \eta \alpha^{3/2} \tilde{\Phi}_*^1 + R_*,
\end{equation}
where $\Phi_y^{(0)}$ is the component of $\Phi_y$ in the zero-th Fock sector and the conditions
\begin{equation}\label{eq:property_R*}
R_*^{(0)} = 0, \qquad    \langle \tilde{\Phi}_*^1\,|\, R_*^{(1)}\rangle_* = 0,
\end{equation}
define $R_*$ and $\eta$.
\item We introduce $\kappa \in \mathbb{C}$ and $R^{\#}_y$, defined by 
\begin{equation}
R_y = \kappa \Phi_{\#}^y + R^{\#}_y,
\end{equation}
and the conditions $\langle R^{\#}_y\,|\, \Phi_{\#}^y\rangle_{\#} = 0$ and $R^{\#(0)}_y = 0$.
\end{itemize}
We observe that, by construction, for all the following vectors holds
\begin{equation}
\Psi_y, u_{\alpha} \otimes \Phi_y, \Phi_{\#}^y, R^{\#}_y \in \mathscr{D}(H_y) \subseteq H_0^1(\mathbb{R}^3_y) \otimes \mathscr{D}(d\Gamma(|k|)).
\end{equation}
We calculate now the quadratic form of $H_y$ on $\Psi_y$.
\begin{equation}\label{eq:quad_lower_expansion}
\langle \Psi_y\,|\,H_y \,|\,\Psi_y\rangle = \langle \,H_y \rangle_{u_{\alpha} \otimes \Phi_y} + \langle \,H_y \rangle_{R_y} + 2 \mathrm{Re} \langle u_{\alpha} \otimes \Phi_y\,|\,R_y\rangle.
\end{equation}
We analyze each term separately. Let us start from the quadratic form in $u_{\alpha} \otimes \Phi_y$:
\begin{align}
\langle \,H_y \rangle_{u_{\alpha} \otimes \Phi_y} &= \langle H_y \rangle_{u_{\alpha}\otimes \Phi_y^{(0)}} + \langle H_y \rangle_{u_{\alpha}\otimes 2\alpha^{3/2}\eta\Phi_*^1}  + \langle H_y \rangle_{u_{\alpha}\otimes R_*} \nonumber \\
&\quad+ 2 \mathrm{Re} \langle u_{\alpha}\otimes \Phi_y^{(0)}\,|\,H_y u_{\alpha} \otimes (2\alpha^{3/2}\eta\Phi_*^1)\rangle \nonumber \\
&\quad+ 2 \mathrm{Re} \langle u_{\alpha}\otimes \Phi_y^{(0)}\,|\,H_y u_{\alpha} \otimes R_*\rangle \nonumber\\
&\quad+ 2 \mathrm{Re} \langle u_{\alpha}\otimes R_*\,|\,H_y (2\alpha^{3/2}\eta\Phi_*^1)\rangle, \label{eq:quad_u}
\end{align}
where
\begin{equation}\label{eq:quad_u_0}
\langle H_y \rangle_{u_{\alpha}\otimes \Phi_y^{(0)}} = \Big(e_{\alpha} + \alpha \|\lambda_y\|^2 -\frac{\alpha^2}{L^3} + O\Big( \frac{\alpha^2}{L^5}\Big) + O\big( \alpha^2 L e^{-L/2}\big) \Big) |\Phi_y^{(0)}|^2 ,
\end{equation}
thanks to \eqref{eq:potential_out_loc}, and
\begin{equation}\label{eq:u_Phi1*}
\langle H_y \rangle_{u_{\alpha}\otimes 2\alpha^{3/2}\eta\Phi_*^1} = 4 \alpha^3|\eta|^2 \|\tilde{\Phi}_*^1\|^2_* + O(\alpha^4) = 4 \alpha^3|\eta|^2 \|\Phi_*^1\|^2_* + O(\alpha^4),
\end{equation}
due to \eqref{eq:V<_prop}, \eqref{eq:V>_prop} and by symmetries in the $*-$norm. The quadratic term for $R_*$ gives
\begin{align}
\langle H_y \rangle_{u_{\alpha}\otimes R_*} &= (e_{\alpha} + \alpha \|\lambda_y\|^2)\|R_*\|^2 + \|R_*\|^2_* + \alpha\langle V_y\rangle_{u_{\alpha}}\|R_*\|^2 +\alpha \|A^-_y R_*\|^2 \nonumber \\
&\geq \|R_*\|^2_* + C_1 \alpha \|R_*\|^2, \label{eq:quad_R*}
\end{align}
where we bounded from below $\|A^-_y R_*\|^2$ by zero, used \eqref{eq:V<_prop}, \eqref{eq:V>_prop} with \eqref{eq:V>u_prop} and chose $ 0 < C_1 \leq \|\lambda_y\|^2 + \alpha^{-1}e_{\alpha} -\frac{\alpha}{L}- C\alpha L e^{-L/2}$. 

 For the cross terms we have 
\begin{equation}\label{eq:u_Phi0_Phi1*}
2 \mathrm{Re} \langle u_{\alpha}\otimes \Phi_y^{(0)}\,|\,H_y (u_{\alpha} \otimes 2\alpha^{3/2}\eta\tilde{\Phi}_*^1)\rangle =  0, 
\end{equation}
because $P_j u_{\alpha} \bot u_{\alpha}, j =1,2,3,$ by mismatch of Fock sectors, and 
\begin{equation}\label{eq:u_Phi0_R*}
2 \mathrm{Re} \langle u_{\alpha}\otimes \Phi_y^{(0)}\,|\,H_y (u_{\alpha} \otimes R_*)\rangle =  0, 
\end{equation}
because $R^{(0)}_* = 0$. We use again this last property  together with \eqref{eq:property_R*} in 
\begin{align}
2 \mathrm{Re} &\langle u_{\alpha}\otimes R_*\,|\,H_y (u_{\alpha} \otimes 2\alpha^{3/2}\eta\tilde{\Phi}_*^1)\rangle\nonumber \\  &=4 (e_{\alpha} + \alpha \|\lambda_y\|^2 +\alpha \langle V_y\rangle_{u_{\alpha}}) \mathrm{Re} \big(\eta\langle  R_*\,|\,\alpha^{3/2} \tilde{\Phi}_*^1\rangle \big) + 2\alpha\mathrm{Re} \langle A^-_y R_*\,|\, A^-_y 2\eta\alpha^{3/2} \tilde{\Phi}_*^1\rangle. \label{eq:middlestep_u_R*_Phi1}
\end{align}
We now apply a Cauchy-Schwarz inequality, for both the scalar products, weighted with a parameter $\varepsilon_1 >0$ to be chosen later and the standard Fock estimate, for any $\Psi \in \Gamma_s(\mathfrak{h}_+)$,
\begin{equation}\label{eq:Fockestimate}
\|A^-_y \Psi\| \leq \|\lambda_y\|_{L^{\infty}(\mathbb{R}^3_y;\mathfrak{h}_+)} \|\mathcal{N}^{1/2}\Psi\|,
\end{equation}
where $\mathcal{N}$ is the number operator, to obtain
\begin{align}
\eqref{eq:middlestep_u_R*_Phi1} \geq& -C \alpha (\alpha^{-1}e_{\alpha} + \alpha \|\lambda_y\|^2 + \alpha^2 L e^{-L/2})  \varepsilon_1 \|R_*\|^2 -C  \alpha^4\varepsilon_1^{-1}|\eta|^2\|\tilde{\Phi}_*^1\|^2 \nonumber\\
\geq& -C_2 \alpha \varepsilon_1 \|R_*\|^2 + O(\alpha^4), \label{eq:u_R*_Phi1*}
\end{align}
where we used also \eqref{eq:V<_prop}, \eqref{eq:V>_prop} and chose $0 < C_2 \leq C(\|\lambda_y\|^2 + \alpha^{-1}e_{\alpha} +  C \alpha L e^{-L/2})$.

This implies that, using \eqref{eq:quad_u_0}, \eqref{eq:u_Phi1*}, \eqref{eq:quad_R*}, \eqref{eq:u_Phi0_Phi1*}, \eqref{eq:u_Phi0_R*} and \eqref{eq:u_R*_Phi1*} in \eqref{eq:quad_u}, we get
\begin{align}
\langle H_y \rangle_{u_{\alpha} \otimes \Phi_y}  &\geq \Big(e_{\alpha} + \alpha \|\lambda_y\|^2 -\frac{\alpha^2}{L^3}\Big) |\Phi_y^{(0)}|^2 +4 \alpha^3|\eta|^2 \|\Phi_*^1\|^2_* + \|R_*\|_*^2 \nonumber\\
&\quad +\alpha(C_1 -C_2\varepsilon_1 )\|R_*\|^2+ O\Big( \frac{\alpha^2}{L^5}\Big) + O\big( \alpha^2 L e^{-L/2}\big) 
 + O(\alpha^4). \label{eq:quad_u_result}
\end{align}
For the quadratic form of the remainder, 
\begin{equation}
\langle H_y\rangle_{R_y} = \langle H_y\rangle_{\kappa \Phi_{\#}^y} + \langle H_y\rangle_{R_y^{\#}} + 2 \mathrm{Re} \langle \kappa\Phi_{\#}^y\,|\, H_y R_y^{\#}\rangle, 
\end{equation}
we have 
\begin{align}
\langle H_y\rangle_{\kappa \Phi_{\#}^y} &= |\kappa|^2 \big( \|\Phi_{\#}^y\|_{\#}^2 + (e_{\alpha}+ \alpha \|\lambda_y\|^2 )\|\Phi_{\#}^y\|^2  +\alpha \|A^-_y \Phi_{\#}^y\|^2 + \alpha \langle V_y\rangle_{ \Phi_{\#}^y}\big) \nonumber\\
&= |\kappa|^2 \|\Phi_{\#}^y\|_{\#}^2 + O(\alpha^4\log(\alpha^{-1})), \label{eq:quad_Phi_ort}
\end{align}
where we used \eqref{eq:norm_Phi_ort}, \eqref{eq:norm_A-Phi_ort} and \eqref{eq:Vy_Phi_ort}.
For the quadratic form in $R^{\#}_y$ we have
\begin{equation}
\langle H_y\rangle_{R^{\#}_y} = 	\|R^{\#}_y\|^2_{\#} + (e_{\alpha} +\alpha \|\lambda_y\|^2 )\|R_y^{\#}\|^2 + \alpha \|A^-_y R^{\#}_y\|^2 + \alpha \langle V_y\rangle_{ R^{\#}_y}. \label{eq:quad_Ry}
\end{equation}
We estimate $\|A^-_y R^{\#}_y\|^2$ by zero for a lower bound, and we observe that 
\begin{align}
\|R^{\#}_y\|^2_{\#} &= \|P R^{\#}_y\|^2 -\Big\langle \frac{\alpha}{|x|} \Big\rangle_{R^{\#}_y} -e_{\alpha}\|R^{\#}_y\|^2 + \|R^{\#}_y\|^2_* \nonumber\\
&\geq  (1- \varepsilon_2^{-1} \alpha) \|P R^{\#}_y\|^2 - (\varepsilon_2 \alpha + e_{\alpha}) \|R^{\#}_y\|^2+ \|R^{\#}_y\|^2_*,
\end{align}
where we used a Hardy-type inequality for the Coulomb potential, a Cauchy-Schwarz weighted with a parameter $\varepsilon_2 >0$, and
\begin{align}
 \langle \alpha V_y\rangle_{R^{\#}_y} \geq& -C\frac{\alpha^2}{L} \|R^{\#}_y\|^2   + \langle \alpha V^>_y\rangle_{R^{\#}_y}  \nonumber\\
\geq& -\Big( C\frac{\alpha^2}{L} + \varepsilon_3 \alpha\Big) \|R^{\#}_y\|^2 - \varepsilon_3^{-1} \alpha \|P R^{\#}_y\|^2,
\end{align}
thanks again to a Cauchy-Schwarz inequality, this time weighted with $\varepsilon_3>0$, and used \eqref{eq:V>_prop}.
We conclude that 
\begin{equation}\label{eq:quad_R_ort}
\langle H_y\rangle_{R^{\#}_y} \geq (1- (\varepsilon_2^{-1}+\varepsilon_3^{-1})  \alpha) \|P R^{\#}_y\|^2 + (C_3- \varepsilon_2 -\varepsilon_3 )\alpha \|R^{\#}_y\|^2+ \|R^{\#}_y\|^2_*, 
\end{equation}
where we chose $0 < C_3 \leq \|\lambda_y\|^2 + \alpha^{-1}e_{\alpha} - C \frac{\alpha}{L}$.

For the cross term, using the orthogonality of $R^{\#}_y$ and $\Phi_{\#}^y$ in the $\#-$scalar product,
\begin{align}
2 \mathrm{Re} \langle \kappa\Phi_{\#}^y\,|\, H_y R_y^{\#}\rangle &= (e_{\alpha} + \alpha \|\lambda_y\|^2)\, 2 \mathrm{Re}\langle \kappa \Phi_{\#}^y\,|\, R^{\#}_y\rangle \nonumber \\
&\quad +  2 \alpha\mathrm{Re} \langle A^-_y \kappa \Phi_{\#}^y\,|\, A^-_y R^{\#}_y\rangle + 2 \alpha \mathrm{Re} \langle \kappa \Phi_{\#}^y\,|\,V_y R^{\#}_y \rangle, \label{eq:Phi_ort_R}
\end{align}
and we observe that, thanks to a Cauchy-Schwarz inequality weighted with a parameter $\varepsilon_4 >0$,
\begin{align}\label{eq:A-_A-_Phi_ort_R}
2 \alpha\mathrm{Re} \langle A^-_y \kappa \Phi_{\#}^y\,|\, A^-_y R^{\#}_y\rangle \geq&   - C \alpha \varepsilon_4 \|A^-_y R^{\#}_y\|^2-C \varepsilon_4^{-1} \alpha |\kappa|^2\|A^-_y\Phi_{\#}^y\|^2,\nonumber \\
\geq&  - C \alpha \varepsilon_4 \| R^{\#}_y\|^2 + O(\alpha^4 ),
\end{align}
where we used Lemma \ref{lem:norm_vector_ort} and \eqref{eq:Fockestimate}.
By a Cauchy-Schwarz inequality, \eqref{eq:V<_prop}, \eqref{eq:V>_prop} and Lemma \ref{lem:norm_vector_ort} we have
\begin{align}
2 \alpha \mathrm{Re} \langle \kappa \Phi_{\#}^y\,|\,V_y R^{\#}_y \rangle \geq& -C \alpha \varepsilon_5^{-1}|\kappa|^2 \|V_y\Phi_{\#}^y\|^2 - C \alpha \varepsilon_5 \|R^{\#}_y\|^2 \nonumber\\
\geq&  \;O(\alpha^4) - C \alpha \varepsilon_5 \|R^{\#}_y\|^2.  \label{eq:Phi_ort_V_R}
\end{align}
 Again by Lemma \ref{lem:norm_vector_ort} and by a Cauchy-Schwarz weighted with a parameter $\varepsilon_6>0$, we have, since $\alpha^{-1}e_{\alpha} + \|\lambda_y\|^2$ is positive,
\begin{align}
(e_{\alpha} + \alpha \|\lambda_y\|^2)\, 2 \mathrm{Re}\langle \kappa \Phi_{\#}^y\,|\, R^{\#}_y\rangle &\geq - C \varepsilon_6 \alpha \|R^{\#}_y\|^2 -C \varepsilon_6^{-1} \alpha |\kappa|^2 \|\Phi_{\#}^y\|^2  \nonumber \\
&=  - C \varepsilon_6 \alpha \|R^{\#}_y\|^2 + O(\alpha^4 \log(\alpha^{-1})). \label{eq:e_phi_ort_R}
\end{align}
Collecting \eqref{eq:A-_A-_Phi_ort_R}, \eqref{eq:Phi_ort_V_R}, \eqref{eq:e_phi_ort_R} and plugging them into \eqref{eq:Phi_ort_R} we get
\begin{equation}
2 \mathrm{Re} \langle \kappa\Phi_{\#}^y\,|\, H_y R_y\rangle  \geq - C (\varepsilon_4 + \varepsilon_5 +\varepsilon_6) \alpha \|R^{\#}_y\|^2  + O(\alpha^4 \log(\alpha^{-1})),
\end{equation} 
and using this last estimate, \eqref{eq:quad_Phi_ort} and \eqref{eq:quad_R_ort} we get
\begin{align}
\langle H_y\rangle_{R_y}  \geq |\kappa|^2 \|\Phi_{\#}^y\|_{\#}^2  + \|R^{\#}_y\|^2_* + \alpha \Big(C_3 - C\sum_{j=2}^6 \varepsilon_j \Big)\|R^{\#}_y\|^2  +(1- C\alpha) \|P R^{\#}_y\|^2+ O(\alpha^4 \log(\alpha^{-1})).\label{eq:quad_R_result}
\end{align}

Now we analyze the last term in \eqref{eq:quad_lower_expansion}
\begin{align}\label{eq:Phi0_Phi1_R}
2 \mathrm{Re} \langle u_{\alpha}\otimes \Phi_y\,|\, H_y R_y \rangle &= 2 \mathrm{Re} \langle u_{\alpha}\otimes \Phi_y^{(0)} \,|\, H_y R_y \rangle + 2 \mathrm{Re} \langle u_{\alpha}\otimes 2 \alpha^{3/2}\eta \Phi_*^1\,|\, H_y R_y \rangle \nonumber \\
 &\quad +2 \mathrm{Re} \langle u_{\alpha}\otimes R_*\,|\, H_y R_y \rangle.  
\end{align}
Since $\Phi_{y}^{(0)}$ is in the zero-th Fock sector we have, reconstructing the vector $\Phi_{\#}^y$,
\begin{equation}\label{eq:u_Phi_R_ort}
2 \mathrm{Re} \langle u_{\alpha}\otimes \Phi_y^{(0)} \,|\, H_y R_y \rangle = 4 \alpha^{1/2}\mathrm{Re} (\overline{\Phi}^{(0)}_y\langle P u_{\alpha}\otimes A^+_y \Omega \,|\,  R_y \rangle) 
= -2 \mathrm{Re} (\overline{\Phi}^{(0)}_y\kappa)\|\Phi_{\#}^y\|^2_{\#},
\end{equation}
where we used Lemma \ref{lem:norm_vector_ort} and that $2 \alpha^{1/2}\langle u_{\alpha}\otimes \Omega \,|\, PA^-_y R_y^{\#} \rangle = \langle \Phi_{\#}^y \,|\, R^{\#}_y\rangle_{\#} = 0$.
For the remaining term
\begin{align}
2 \mathrm{Re} \langle u_{\alpha}\otimes 2 \alpha^{3/2}\eta \Phi_*^1\,|\, H_y R_y \rangle &= 2 \mathrm{Re} \langle u_{\alpha}\otimes 2 \alpha^{3/2}\eta \tilde{\Phi}_*^1\,|\, \alpha V_y \Phi^y_{\#} \rangle \nonumber \\
&\quad + 2 \mathrm{Re} \langle u_{\alpha}\otimes 2 \alpha^{3/2}\eta \tilde{\Phi}_*^1\,|\, \alpha V_y R^{\#}_y \rangle, \label{eq:Phi*_R_ort_sum}
\end{align}
because $u_{\alpha} \bot R_y^{\#}$.
For the first addend we use a Cauchy-Schwarz, \eqref{eq:V<_prop}, \eqref{eq:V>_prop} and Lemma \ref{lem:norm_vector_ort} to get
\begin{equation}\label{eq:Phi*_R_ort_Phi}
2 \mathrm{Re} \langle u_{\alpha}\otimes 2 \alpha^{3/2}\eta \tilde{\Phi}_*^1\,|\, \alpha V_y \Phi^y_{\#} \rangle \geq -C |\eta|^2 \alpha^4 \|\tilde{\Phi}_*^1\|^2 -C\alpha |\kappa|^2 \|V_y \Phi_{\#}^y\|^2 = O(\alpha^4).
\end{equation}
For the second addend we use a Cauchy-Schwarz weighted with a parameter $\varepsilon_7>0$, \eqref{eq:V<_prop}, \eqref{eq:V>_prop} to get
\begin{align}
2 \mathrm{Re} &\langle u_{\alpha}\otimes 2 \alpha^{3/2}\eta \tilde{\Phi}_*^1\,|\, \alpha V_y R^{\#}_y \rangle  \nonumber\\
\geq&-C |\eta|^2 \alpha^4 \varepsilon_7^{-1}\|\tilde{\Phi}_*^1\|^2 - C\alpha \varepsilon_7 \Big( \frac{\alpha}{L}\|R^{\#}_y\|^2 + \|P R^{\#}_y\|^2\Big) \nonumber \\
\geq& \;O(\alpha^4) - C \varepsilon_7\alpha \|R^{\#}_y\|^2 - C\alpha \varepsilon_7 \|P R^{\#}_y\|^2.
 \label{eq:Phi*_R_ort}
\end{align}

We now turn the attention to 
\begin{align}
2 \mathrm{Re} \langle u_{\alpha}\otimes R_*\,|\, H_y R_y \rangle &=  2 \alpha\mathrm{Re} \langle u_{\alpha}\otimes R_*\,|\, V_y R_y \rangle \nonumber \\
&=  2 \alpha\mathrm{Re} \langle u_{\alpha}\otimes R_*\,|\, V_y \Phi_{\#}^y \rangle + 2 \alpha\mathrm{Re} \langle u_{\alpha}\otimes R_*\,|\, V_y R_y^{\#} \rangle,
\end{align}
where we used that $u_{\alpha} \bot R_y$.
For the first addend we use a Cauchy-Schwarz inequality weighted with a parameter $\varepsilon_8>0$ and Lemma \ref{lem:norm_vector_ort} to get
\begin{align}
2 \alpha\mathrm{Re} \langle u_{\alpha}\otimes R_*\,|\, V_y \Phi_{\#}^y \rangle &\geq - C \alpha \varepsilon_8 \|R_*\|^2 - C \alpha \varepsilon_8^{-1} |\kappa|^2 \|V_y\Phi_{\#}^y\|^2 \nonumber\\
&= - C \alpha \varepsilon_8 \|R_*\|^2 + O(\alpha^4).\label{eq:R*_u_Phi_ort}
\end{align}
For the second addend we apply a Cauchy-Schwarz inequality and use \eqref{eq:V>u_prop} and \eqref{eq:V<_prop} to get 
\begin{align}
2 \alpha\mathrm{Re} \langle u_{\alpha}\otimes R_*\,|\, V_y R_y^{\#} \rangle \geq& -C\alpha \|V_y u_{\alpha}\| \|R_*\| \|R^{\#}_y\| \nonumber \\
\geq& - C\alpha^2 \|R_*\|^2 - C \alpha^2 \|R^{\#}_y\|^2. \label{eq:R*_u_R_ort}
\end{align}

Collecting then \eqref{eq:u_Phi_R_ort}, \eqref{eq:Phi*_R_ort_Phi}, \eqref{eq:Phi*_R_ort},  \eqref{eq:R*_u_Phi_ort} and \eqref{eq:R*_u_R_ort} and plugging them into \eqref{eq:Phi0_Phi1_R} we get
\begin{align}
2 \mathrm{Re} \langle u_{\alpha}\otimes \Phi_y\,|\, H_y R_y \rangle =& -2 \mathrm{Re} (\overline{\Phi}^{(0)}_y\kappa)\|\Phi_{\#}^y\|^2_{\#} -C \alpha(\varepsilon_7 + \alpha ) \|R_y^{\#}\|^2  \nonumber\\
&  -C \alpha (\varepsilon_8 + \alpha)  \|R_*\|^2 - C\alpha \varepsilon_7 \|P R^{\#}_y\|^2+ O(\alpha^4).\label{eq:Phi0_Phi1_R_result}
\end{align}

We finally collect the inequalities \eqref{eq:quad_u_result}, \eqref{eq:quad_R_result} and  \eqref{eq:Phi0_Phi1_R_result} and plug them into \eqref{eq:quad_lower_expansion} to obtain the following lower bound for the quadratic form of the Hamiltonian
\begin{align*}
\langle \Psi_y\,|\, H_y\,|\,\Psi_y \rangle 
&\geq  \Big(e_{\alpha} + \alpha \|\lambda_y\|^2 -\frac{\alpha^2}{L^3}\Big) |\Phi_y^{(0)}|^2 +\big(|\kappa|^2-2 \mathrm{Re} (\overline{\Phi}^{(0)}_y\kappa)\big) \|\Phi_{\#}^y\|^2_{\#}+4 \alpha^3|\eta|^2 \|\Phi_*^1\|^2_*\\
&\quad + \|R_*\|_*^2 + \alpha(C_1 - C_2 \varepsilon_1 - C\varepsilon_8-C\alpha )\|R_*\|^2  + \alpha \Big(C_3 -\alpha -\sum_{j=2}^7 \varepsilon_j \Big)\|R^{\#}_y\|^2  \\
  &\quad + (1  - C\alpha (1+\varepsilon_7)) \|P R^{\#}_y\|^2    + O\Big( \frac{\alpha^2}{L^5}\Big) + O\big( \alpha^2 L e^{-L/2}\big) + O(\alpha^4 \log(\alpha^{-1})).
\end{align*}
Choosing the $\varepsilon_j$, $j=1,...,8$ such that 
\begin{align*}
C_1 - C_2 \varepsilon_1 - C\varepsilon_8-C\alpha >0, \qquad
C_3 -\alpha -\sum_{j=2}^7 \varepsilon_j >0, \qquad
1 - C \alpha - C\alpha \varepsilon_7 >0.
\end{align*}
we can bound from below the positive terms involving $\|R_*\|^2$, $\|R^{\#}_y\|^2$, $\|P R^{\#}_y\|^2$ and $\|R_*\|^2_*$.
Using that $|\Phi_y^{(0)}|\leq 1$, we complete the square and bound
\begin{equation}
|\kappa|^2-2 \mathrm{Re} (\overline{\Phi}^{(0)}_y\kappa) = |\kappa - \Phi^{(0)}_y|^2 - |\Phi^{(0)}_y|^2 \geq   -1,
\end{equation}
finally making us obtain
\begin{align*}
\langle \Psi_y\,|\, H_y\,|\,\Psi_y \rangle &\geq  \Big(e_{\alpha} + \alpha \|\lambda_y\|^2 -\frac{\alpha^2}{L^3}\Big) |\Phi_y^{(0)}|^2 - \|\Phi_{\#}^y\|^2_{\#}+4 \alpha^3|\eta|^2 \|\Phi_*^1\|^2_*\\
  &\quad + O\Big( \frac{\alpha^2}{L^5}\Big) + O\big( \alpha^2 L e^{-L/2}\big) + O(\alpha^4 \log(\alpha^{-1})).
\end{align*}
Comparing the result with the upper bound obtained in Theorem \ref{thm:upper_bound} let us bound $|\Phi^{(0)}_y|^2$ and $|\eta|^2$ by $1$ plus terms which, multiplied with the rest, can be reabsorbed in the error terms, concluding the proof of the desired lower bound.
\end{proof}

\subsection{Evaluation of the norms}\label{subsec:norms_evaluation}

Joining together the upper and lower bounds for $E_y$ obtained in Theorem \ref{thm:upper_bound} and Theorem \ref{thm:lower_bound} and subtracting the estimate of $E_{\infty}$ by Theorem \ref{thm:freebound}, we get
\begin{align}
W_y^{QFT} = E_y - E_{\infty} &=  -\frac{\alpha^2}{L^3} + \alpha ( \| \lambda_y\|^2 -  \|\lambda_{\infty}\|^2) + \|\Phi_{\#}^{\infty}\|^2_{\#}- \|\Phi_{\#}^y\|^2_{\#}  \\
&\quad+ O\Big( \frac{\alpha^2}{L^5}\Big) + O\big( \alpha^2 L e^{-L/2}\big) + O(\alpha^4 \log(\alpha^{-1})).
\end{align}
Introducing the quantity
\begin{equation}
\mathscr{E} := \alpha ( \| \lambda_y\|^2 -  \|\lambda_{\infty}\|^2) + \|\Phi_{\#}^{\infty}\|^2_{\#}- \|\Phi_{\#}^y\|^2_{\#},
\end{equation}
our goal in this section is to prove an estimate for $\mathscr{E}$ by the following proposition.
\begin{proposition}\label{propos:E_norm_estimates}
For any $L>1$,
\begin{equation}
\mathscr{E} = \frac{\alpha^2}{L^3} -\frac{\alpha}{L^4}\aleph_{\alpha} + O \Big( \frac{\alpha^3}{L^2}\log(\alpha^{-1})\Big)  + O(\alpha^4 \log(\alpha^{-1})) + O(\alpha^3 L^2 e^{-L}).
\end{equation}
\end{proposition}

\begin{proof}
We split the proof in three parts: we evaluate the norms involving the $\lambda$ terms, then the ones involving the $\Phi$ terms and finally we sum the results and give the estimate above by studying some complex line integrals.

Let us recall the definitions \eqref{eq:lambda_0_def} and \eqref{eq:lambda_y_0_def} of $\lambda_y$ and $\lambda_{\infty}$, respectively, for reader's convenience.
Then, 
\begin{align*}
\|\lambda_{y}\|^2 =  \int_{\mathbb{R}^3_+ } \mathrm{d}k  \,\frac{\chi^2_{\Lambda}(k)}{\pi^2|k| }
\sum_{\gamma=1,2} \left\{ \mathbf{e}^{(1)\,2}_{\gamma}(k) \cos^2(k_1 y) + (\mathbf{e}^{(2)\,2}_{\gamma}(k) +
\mathbf{e}^{(3)\,2}_{\gamma}(k)) \sin^2(k_1 y) \right\},
\end{align*}
where we denoted by $\mathbb{R}^3_+ := \mathbb{R}^+ \times \mathbb{R}^2.$
We now use that $\{\hat{k}, \mathbf{e}_{1}(k), \mathbf{e}_2(k) \}$ is an orthonormal basis, for a.e. $k \in \mathbb{R}^3$, to have
\begin{equation}
\sum_{\gamma=1,2} \mathbf{e}_{\gamma}^{(j)2} = 1 - \hat{k}_j^2, \qquad j= 1,2,3,
\end{equation}
and plug in the previous expression, using some goniometric formulas, to get
\begin{align*}
\|\lambda_y\|^2= \int_{\mathbb{R}^3_+ } \mathrm{d}k\,  \frac{\chi^2_{\Lambda}(k) }{2 \pi^2|k| } \left\{ (1-\hat{k}_1^2) (1 +   \cos(2k_1 y)) + (2-\hat{k}_2^2 - \hat{k}_3^2)   (1  - \cos(2k_1y)) \right\}. 
\end{align*}
Since we have only even integrands in the $k_1$ variable, we can turn the integration on the whole $\mathbb{R}^3$ getting a factor $1/2$, and we separate also the integer from the oscillatory parts:
\begin{equation*}
=\int_{\mathbb{R}^3 } \mathrm{d}k\, \frac{\chi^2_{\Lambda}(k)}{4\pi^2|k| }(3 -\hat{k}_1^2 -\hat{k}_2^2-\hat{k}_3^2 )+\int_{ \mathbb{R}^3} \mathrm{d}k \frac{\chi^2_{\Lambda}(k)}{4\pi^2|k|} (-1-\hat{k}_1^2+\hat{k}_2^2+\hat{k}_3^2)\cos(2k_1 y). 
\end{equation*}
Using that $3 -\hat{k}_1^2 -\hat{k}_2^2-\hat{k}_3^2 = 3-|\hat{k}|^2= 2$, we recognize the first term to be the expression of the norm for $\lambda_{\infty}$ and therefore we obtain
\begin{equation}\label{eq:result_lambda_norm}
\alpha(\|\lambda_y\|^2 -\|\lambda_{\infty}\|^2)  =\alpha \int_{\mathbb{R}^3 }\mathrm{d}k \, f_y(k),
\end{equation}
where we denoted by 
\begin{equation}\label{eq:fexpression}
f_y(k) := \frac{1}{4 \pi^2}\frac{\chi^2_{\Lambda}(k)}{|k|} (-1-\hat{k}_1^2+\hat{k}_2^2+\hat{k}_3^2) \cos(2k_1y).
\end{equation}

We turn now the attention to the $\Phi$ terms. Let us  calculate
\begin{align*}
\|\Phi_{\#}^y\|^2_{\#} &= 4 \alpha \|(h_{\alpha} - e_{\alpha} +H_f^+)^{-1/2} Pu_{\alpha} \otimes A_{y}^+\Omega\|^2 \\
&=4 \alpha \int_{\mathbb{R}^3_+\times \mathbb{R}^3} \mathrm{d}k\, \mathrm{d}x \; \left| \frac{Pu_{\alpha}}{(h_{\alpha}-e_{\alpha}+|k|)^{1/2}} \lambda_y(k) \right|^2+O(\alpha^3 L^2 e^{-L}).
\end{align*}
where the last error term was obtained completing the domain of integration in the position variable and using that 
\begin{equation}
\int_{\{x_1 \geq y\}} dx \,| (h_{\alpha}-e_{\alpha} + |k|)^{-1/2} P u_{\alpha}(x) |^2\leq \frac{C}{|k|} \alpha^3 L^2 e^{-L}.
\end{equation}
Expanding the square, we obtain, by similar calculations to the $\lambda$ terms and recalling the expression of $f_y(k)$ in \eqref{eq:fexpression},
\begin{multline*}
\|\Phi_{\#}^y\|^2_{\#} = 8 \alpha \int_{\mathbb{R}^3_+\times \mathbb{R}^3} \mathrm{d}k \mathrm{d}x \; \left| \frac{P u_{\alpha}}{(h_{\alpha}-e_{\alpha}+|k|)^{1/2}} \lambda_{\infty}(k) \right|^2 +\\
+ 8 \alpha \int_{\mathbb{R}^3_+\times \mathbb{R}^3} \mathrm{d}k \mathrm{d}x \,  \left| \frac{P u_{\alpha}}{(h_{\alpha}-e_{\alpha}+|k|)^{1/2}}\right|^2  f_y(k)+O(\alpha^3 L^2 e^{-L}).
\end{multline*} 
We observe that both the integrals can be extended to the whole $\mathbb{R}^3$ dropping a factor $2$ thanks to the even integrand, recovering the expression for $\|\Phi_{\#}^{\infty}\|_{\#}^2$ and therefore write, 
\begin{equation}
\|\Phi_{\#}^{\infty}\|^2_{\#} - \|\Phi_{\#}^y\|^2_{\#} = - 4 \alpha \int_{\mathbb{R}^3}\mathrm{d}k \;   \left\| \frac{P u_{\alpha}}{(h_{\alpha}-e_{\alpha}+|k|)^{1/2}}\right\|_{L^2}^2 f_y(k) +O(\alpha^3 L^2 e^{-L}).
\end{equation}
We use the relation $2 i Pu_{\alpha} =  (h_{\alpha}-e_{\alpha}) x u_{\alpha}$ to write
\begin{equation}
\|\Phi_{\#}^{\infty}\|^2_{\#} - \|\Phi_{\#}^y\|^2_{\#} = -   \alpha \int_{\mathbb{R}^3}\mathrm{d}k \;   \left\| \frac{(h_{\alpha}-e_{\alpha}) x u_{\alpha}}{(h_{\alpha}-e_{\alpha}+|k|)^{1/2}}\right\|_{L^2}^2 f_y(k) +O(\alpha^3 L^2 e^{-L}) .
\end{equation}
Calling 
\begin{equation}
G(k) :=\left\| \frac{(h_{\alpha}-e_{\alpha}) x u_{\alpha}}{(h_{\alpha}-e_{\alpha}+|k|)^{1/2}}\right\|_{L^2}^2,
\end{equation}
we can finally write
\begin{equation}\label{eq:result_Phi_norm}
\|\Phi_{\#}^{\infty}\|^2_{\#} - \|\Phi_{\#}^y\|^2_{\#}= - \alpha \int_{\mathbb{R}^3}\mathrm{d}k \;  G(k)\,f_y(k)+O(\alpha^3 L^2 e^{-L}).
\end{equation}

Let us calculate now the difference between the norms of the $\lambda$ and $\Phi$ terms. In order to do so, we observe that, by explicit calculations,
\begin{equation}\label{eq:relationhydmom}
\langle x u_{\alpha} \,|\,(h_{\alpha}-e_{\alpha})\,|\, x u_{\alpha}\rangle = 3.
\end{equation}
Therefore, by \eqref{eq:relationhydmom}, \eqref{eq:result_lambda_norm} and \eqref{eq:result_Phi_norm}, 
\begin{align*}
\mathscr{E} =& \alpha \int_{\mathbb{R}^3} \mathrm{d} k \; f_y(k) \, (1- G(k))+O(\alpha^3 L^2 e^{-L}) = \\
=& \alpha \int_{\mathbb{R}^3} \mathrm{d} k \; f_y(k) \, \left\langle x u_{\alpha}\,\bigg|\, \frac{(h_{\alpha}-e_{\alpha})}{3}- 	\, \frac{(h_{\alpha}-e_{\alpha})^2}{(h_{\alpha}-e_{\alpha}+|k|)}\,\bigg|\, x u_{\alpha} \right\rangle +O(\alpha^3 L^2 e^{-L})=\\
=&  \alpha \int_{\mathbb{R}^3} \mathrm{d} k \; f_y(k) \, \left(\left\langle x u_{\alpha}\,\bigg|\, \frac{(h_{\alpha}-e_{\alpha})|k|}{3(h_{\alpha}-e_{\alpha}+|k|)}\,\bigg|\, x u_{\alpha} \right\rangle \right. \\
 &\left.\qquad \qquad \qquad- \left\langle x u_{\alpha}\,\bigg|\, \frac{2(h_{\alpha}-e_{\alpha})^2}{3(h_{\alpha}-e_{\alpha}+|k|)}\,\bigg|\, x u_{\alpha} \right\rangle \right)+O(\alpha^3 L^2 e^{-L}), 
\end{align*}
where we used the second resolvent formula to reduce to a common denominator and perform the calculation above.

The estimate of the oscillatory integrals are proven in  Appendix 2. We show in Lemma \ref{lem:errorhsquare} how the second term in the expression above produces an error of order $O(\frac{\alpha^4}{L})$, while in Proposition \ref{prop:cancellation_VdW} we show how the first integral is responsible for the cancellation of the van der Waals term coming from the Coulomb interaction and produces the new leading term. This concludes the proof of Proposition \ref{propos:E_norm_estimates}.
\end{proof}

This concludes the proof of the main Theorem \ref{thm:main}: by Theorem \ref{thm:upper_bound}, Theorem \ref{thm:lower_bound} and Proposition \ref{propos:E_norm_estimates} we get, observing that the error $O(\alpha^3 L^2 e^{-L})$ can be absorbed in $ O\big( \alpha^2 L e^{-L/2}\big)$,
\begin{align*}
W_L^{\text{QFT}} &= \mathscr{E} - \frac{\alpha^2}{L^3} + O\Big(\frac{\alpha^2}{L^5} \Big) + O(\alpha^4 \log(\alpha^{-1})) + O\big( \alpha^2 L e^{-L/2}\big)= \\
&=  - \frac{\alpha^2}{L^4}\aleph_{\alpha,L} + O\Big(\frac{\alpha^2}{L^5} \Big) +  O\Big(\frac{\alpha^3}{L^2} \log(\alpha^{-1})\Big) + O(\alpha^4\log(\alpha^{-1})) + O\big( \alpha^2 L e^{-L/2}\big),
\end{align*}
with $\aleph_{\alpha,L}$ defined in \eqref{eq:aleph_def}.

\section{Discussion of the result}\label{sec:discussion_result}

In this section we analyze the result of Theorem \ref{thm:main} to get information about the leading term of the interaction energy in the different regimes for the distance.

Let us recall the expression of $\aleph_{\alpha,L}$ obtained in the calculations for the proof of Proposition \ref{prop:cancellation_VdW}:
\begin{equation}
\aleph_{\alpha,L} =  \frac{1}{6 \pi } \left\langle\alpha L\arctan\left(\frac{1}{\alpha L (h_1 - e_1)}\right) \right\rangle_{x u_1}, 
\end{equation}
and the expression of the interaction energy $W_L^{\text{QFT}}$ obtained in Theorem \ref{thm:main} for the reader's convenience
\begin{align}\label{eq:interenergy_expression_evaluation}
W_L^{\text{QFT}} = E_y -E_{\infty}=& -\aleph_{\alpha,L} \frac{\alpha}{L^4}+  O(\alpha^4 \log(\alpha^{-1})) \nonumber \\
&+ O\big( \alpha^2 L e^{-L/2}\big) + O \Big( \frac{\alpha^2}{L^5}\Big) + O\Big( \frac{\alpha^3}{L^2}\log(\alpha^{-1})\Big).
\end{align}
We want to study this expression and compare the first term with the behavior of the error terms. 

The spectral gap value for the spectrum of the hydrogen atom Hamiltonian $h_1$ is $\frac{3}{16}$, so that $(h_1-e_1)^{-1} \leq \frac{16}{3}$.

Let us consider the regime $L \geq \frac{16}{3}\alpha^{-1}$: in this case, the argument of the $\arctan$ in $\aleph_{\alpha,L}$ is smaller than $1$ and by a Taylor expansion and functional calculus it can be approximated by 
\begin{equation}
\aleph_{\alpha,L}  = \frac{1}{6\pi} \|(h_1-e_1)^{-1/2}x u_1\|^2 + O\Big(\frac{1}{\alpha^2 L^2} \Big),
\end{equation}
which, plugged in \eqref{eq:interenergy_expression_evaluation}, it gives that the first two order terms are, since $L \geq \frac{16}{3}\alpha^{-1}$,
\begin{equation}
W_L^{\text{QFT}} \simeq -\frac{\alpha}{6\pi L^4} \|(h_1-e_1)^{-1/2}x u_1\|^2 + O(\alpha^4 \log(\alpha^{-1})), 
\end{equation}
where the second term is dominant and expresses an error bigger than the first term.

For the regime $\frac{16}{3} < L <\frac{16}{3} \alpha^{-1}$, we introduce the parameter $\eta>0$ such that the interval can be described as
\begin{equation}
L = \frac{16}{3} \alpha^{-1 + \eta}, \qquad \eta \in (0,1).
\end{equation}
For these distances, by a Taylor expansion and recalling that $\|x u_1\|^2 = 12$, we have
\begin{equation}
\aleph_{\alpha,L} = \frac{8\alpha^{\eta}}{9\pi} \Big\langle\arctan \left(\frac{3\alpha^{-\eta}}{16 (h_1-e_1)}\right)\Big\rangle_{x u_1} =  \frac{16}{3}\alpha^{\eta} + O(\alpha^{2 \eta}),
\end{equation}
which, plugged in \eqref{eq:interenergy_expression_evaluation}, gives that the relevant terms in the interaction energy are
\begin{equation}
W_L^{\text{QFT}} \simeq - \frac{16}{3}\frac{\alpha^{1+\eta}}{L^4}+ O\Big( \frac{\alpha^{1+2\eta}}{L^4}\Big)  + O(\alpha^4 \log(\alpha^{-1}))+ O (\alpha^{5-2 \eta}).
\end{equation}
The first term is the leading term for $\eta \in \big(\frac{1}{3},1\big)$, otherwise the leading term is of order $O(\alpha^4\log(\alpha^{-1}))$.

For the remaining regime, $1 < L \leq  \frac{16}{3}$, we recover the expression of the van der Waals term, because, again by a Taylor expansion, we have
\begin{equation}
\aleph_{\alpha,L} = \frac{\alpha L}{12} \|x u_1\|^2 + O(\alpha^2 L) = \alpha L +O(\alpha^2 L),
\end{equation}
and then the leading term reads
\begin{equation}
W_{L}^{\text{QFT}} \simeq - \frac{\alpha^2}{L^3}. 
\end{equation}

Finally, we can collect here below the expressions of the leading terms of the energy and the associated values of $\aleph_{\alpha,L}$ depending on the distance:

\begin{equation*}
  \aleph_{\alpha,L} \simeq \begin{dcases}
        \alpha L \quad &\text{if} \quad 1< L \leq \frac{16}{3}, \\
        \frac{16}{3}\alpha^{\eta} \quad &\text{if} \quad L = \frac{16}{3} \alpha^{-1+\eta}, \; \eta \in (0,1),\\
       \frac{1}{6\pi}\|(h_1-e_1)^{-1/2}x u_1\|^2 \quad &\text{if} \quad L \geq \frac{16}{3} \alpha^{-1},
        \end{dcases}
 \end{equation*}
 and
 \begin{equation*}
  W_{L}^{\text{QFT}} \simeq \begin{dcases}
        -\frac{\alpha^2}{L^3} \quad &\text{if} \quad 1< L \leq \frac{16}{3}, \\
        -\frac{16}{3}\frac{\alpha^{1+ \eta}}{L^4} \quad &\text{if} \quad L = \frac{16}{3} \alpha^{-1+\eta}, \; \eta \in \Big(\frac{1}{3},1\Big),\\
       O(\alpha^4\log(\alpha^{-1})) \quad &\text{if} \quad L \geq \frac{16}{3} \alpha^{-1+\frac{1}{3}}.
        \end{dcases}
 \end{equation*}
As a remark, we underline the fact that the expression of the leading term we would have liked to obtain for $W_L^{\text{QFT}}$ in the regime $L > \frac{16}{3}\alpha^{-1}$ is 
\begin{equation}\label{eq:goal_leadterm}
W_{L}^{\text{QFT}} \simeq - \frac{\alpha}{6 \pi L^4}\|(h_1-e_1)^{-1/2} x u_1\|^2,
\end{equation}
but the precision used in the calculation does not allow to produce an error small enough to make the term above to appear as leading term. The problem seems to be, anyway, just of technical nature. Furthermore, it is really intrinsic in the method used that some terms of the error obtained are uniform in $L$. Therefore, whatever the degree of precision of the error in $\alpha$, one can always find a distance large enough such that \eqref{eq:goal_leadterm} is no longer the leading term. 

In the right units, $\alpha^{-1}$ corresponds to the value of a half of Bohr radius, and expressing the distance $y = L\alpha^{-1}$, $L$ represents half of the number of Bohr radii.
In conclusion, plugging the numerical values of the parameters
\begin{equation}
\frac{16}{3} \simeq 5.3, \qquad \frac{16}{3} \alpha^{-1+\frac{1}{3}} \simeq 165, 
\end{equation}
 we see that our result proves the Casimir-Polder effect for all the distances up to approximately $82.5$ Bohr radii.

\section*{Appendix 1: Technical inequalities}\label{app:technical_Phi}

\begin{lemma}\label{lem:norm_vector_ort}
The following estimates hold for the vector $\Phi_{\#}^y$
\begin{align}
\|\Phi_{\#}^y\|^2 &= O(\alpha^3\log(\alpha^{-1})), \label{eq:norm_Phi_ort}\\
\|A^-_y\Phi_{\#}^y\|^2 &= O(\alpha^3), \label{eq:norm_A-Phi_ort}\\
\|P \Phi_{\#}^y\|^2 &= O(\alpha^5 \log(\alpha^{-1})), \label{eq:norm_P_PHi_ort}\\
\langle \alpha V_y\rangle_{\Phi_{\#}^y} &= O(\alpha^5 \log(\alpha^{-1})) \label{eq:Vy_Phi_ort},\\
\|V_y \Phi_{\#}^y\|^2 &= O(\alpha^5 \log (\alpha^{-1})). \label{eq:norm_Vy_Phi_ort}
\end{align}
\end{lemma}
\begin{proof}
Let us start by proving \eqref{eq:norm_Phi_ort}:
\begin{align*}
\|\Phi_{\#}^y\|^2 &= 4 \alpha \|(h_{\alpha} -e_{\alpha}+H_f^+)^{-1} P u_{\alpha} \otimes A^+_y \Omega\|^2 \\
&\leq C \alpha\sum_{\gamma = 1,2}\int_{\mathbb{R}^3_y}\int_{\mathbb{R}^3_+} \diff x\, \diff k\, \frac{\chi_{\Lambda}^2(k)}{|k|}\left|\frac{Pu_{\alpha}}{(h_{\alpha}-e_{\alpha} + |k|)}\right|^2 \\
&\quad \times \big(e^{(1)2}_{\gamma}\cos^2(k_1 y)+ (e^{(2)2}_{\gamma}+e^{(3)2}_{\gamma})\sin^2(k_1 y)  \big)\\
&\leq C \alpha^3  \int_{\mathbb{R}^3}  \diff k \frac{\chi_{\Lambda}^2(k)}{|k|(|k| + \frac{3}{16}\alpha^2)^2} = O(\alpha^3 \log(\alpha^{-1})),
\end{align*}
where we used that the integrand is an even function of $k_1$ to extend the integral to the whole space and that the spectral gap of the hydrogen atom is $\frac{3}{16}\alpha^2$.

For inequality \eqref{eq:norm_A-Phi_ort}, we find, by similar calculations and use of symmetries, that 
\begin{align*}
\|A^-_y\Phi_{\#}^y\| 
&\leq C \alpha \int_{\mathbb{R}^3_y}\diff x\Big(\int_{\mathbb{R}^3_+} \, \diff k\, \frac{\chi_{\Lambda}^2(k)}{|k|}\frac{Pu_{\alpha}}{(h_{\alpha}-e_{\alpha} + |k|)} \\
&\quad \times \sum_{\gamma = 1,2}\big(e^{(1)2}_{\gamma}\cos^2(k_1 y)+ (e^{(2)2}_{\gamma}+e^{(3)2}_{\gamma})\sin^2(k_1 y)  \big)\Big)^2 \\
&\leq C \alpha \int_{\mathbb{R}^3}  \diff x  ((h_{\alpha}-e_{\alpha})x u_{\alpha}(x))^2 \Big( \int_{\mathbb{R}^3}\diff k \frac{\chi_{\Lambda}(k)^2}{|k|^2}\Big)^2 = O(\alpha^3).
\end{align*}

In order to prove \eqref{eq:norm_P_PHi_ort} we observe that $P u_{\alpha}$ is odd. On the subspace of antisymmetric functions the infimum of the spectrum of $h_{\alpha}$ is strictly bigger than $e_{\alpha}$. This, and the fact that we can choose a $\gamma_0>0$ small enough such that $h_{\alpha} + \gamma_0 (-\Delta)$ is a perturbation of $h_{\alpha}$, gives us the inequality $-(1-\gamma_0)\Delta -\frac{\alpha}{|x|} \geq e_{\alpha}$, which implies $P^2 < \gamma_0^{-1}(h_{\alpha} -e_{\alpha})$. We apply this to 
\begin{align*}
\|P\Phi_{\#}^y\|^2 &\leq \gamma^{-1}_0 C \alpha  \int_{ \mathbb{R}^3}  \diff k  \left|\int_{\mathbb{R}^3} dx\,  \frac{\chi_{\Lambda}^2(k) (h_{\alpha}-e_{\alpha})^{1/2} P u_{\alpha}}{|k|(h_{\alpha}-e_{\alpha} + |k|)^2} \right|^2 \\
&\leq \gamma^{-1}_0 C \alpha^5  \int_{ \mathbb{R}^3}  \diff k  \left|\int_{\mathbb{R}^3} dx\,  \frac{\chi_{\Lambda}^2(k) (h_{1}-e_{1})^{1/2} P u_{1}}{|k|(\frac{3}{16}\alpha^2 + |k|)^2} \right|^2  =    O(\alpha^5 \log(\alpha^{-1})),
\end{align*}
where we used similar calculations as in the proof of \eqref{eq:norm_Phi_ort}.

We prove now \eqref{eq:Vy_Phi_ort}: by \eqref{eq:V<_prop}, \eqref{eq:V>_prop}, \eqref{eq:norm_Phi_ort} and \eqref{eq:norm_P_PHi_ort} we have
\begin{align}
|\alpha\langle V_y\rangle_{ \Phi_{\#}^{y}} |&\leq C\frac{\alpha}{y} \|\Phi^y_{\#}\|^2 + \langle \alpha V^>_y \rangle_{ \Phi_{\#}^y} \nonumber\\
&\leq C \frac{\alpha^5}{L} \log(\alpha^{-1}) + \alpha \|\Phi_{\#}^y\| \|P \Phi_{\#}^y\| = O(\alpha^5 \log(\alpha^{-1})).
\end{align}

For inequality \eqref{eq:norm_Vy_Phi_ort} we use again \eqref{eq:V<_prop}, \eqref{eq:V>_prop}, \eqref{eq:norm_Phi_ort} and \eqref{eq:norm_P_PHi_ort}: 
\begin{equation}
\|V_y \Phi_{\#}^2\|^2 \leq C \frac{\alpha^4}{L^2}\log(\alpha^{-1}) + C\|P \Phi_{\#}^y\|^2 = O(\alpha^5 \log(\alpha^{-1})).
\end{equation}
\end{proof}

\section*{Appendix 2: Estimates of oscillatory integrals}\label{app:oscill}

In this technical Appendix we collected the lemmas which prove the estimates of the oscillatory integrals needed to evaluate the quantity $\mathscr{E}$ in Subsection \ref{subsec:norms_evaluation}. 
\begin{lemma}\label{lem:errorhsquare}
For any $L>1$,
\begin{equation}
\alpha \int_{\mathbb{R}^3} \mathrm{d} k \; f_y(k) \, \bigg\langle x u_{\alpha}\,\bigg|\, \frac{(h_{\alpha}-e_{\alpha})^2}{(h_{\alpha}-e_{\alpha}+|k|)}\,\bigg|\, x u_{\alpha} \bigg\rangle =  O\Big( \frac{\alpha^4}{L}\Big).
\end{equation}
\end{lemma}

\bigskip

\begin{proof}
Let us first bound the quantity
\begin{equation}
J:=\int_{\mathbb{R}^3} \mathrm{d} k \,\frac{ f_y(k) }{\alpha^2 + |k|}, 
\end{equation}
where we recall the expression
\begin{equation}
f_y(k) = \frac{1}{4 \pi^2}\frac{\chi^2_{\Lambda}(k)}{|k|} (-1-\hat{k}_1^2+\hat{k}_2^2+\hat{k}_3^2)\cos(2k_1y).
\end{equation}
Let us change to spherical coordinates $(\rho, \varphi, \theta) \in (0,+\infty) \times (0,2\pi) \times (0,\pi)$ so that $\rho = |k|$, $k_1 = \rho \cos(\theta)$ and $(-1-\hat{k}_1^2 + \hat{k}_2^2+ \hat{k}_3^2) = - 2 \cos^2\theta$ and then $J$ gives
\begin{multline*}
 -\frac{1}{2\pi^2}\int_0^{+\infty}\mathrm{d}\rho \int_{0}^{2\pi} \mathrm{d}\varphi \int_0^{\pi}  \mathrm{d}\theta\, \rho  \sin\theta\, \chi_{\Lambda}^2(\rho)\,   \frac{1}{(\alpha^2+\rho)}(\cos\theta)^2 \cos(2\rho y\cos\theta ) \\
=-\frac{1}{\pi}\int_0^{+\infty}\mathrm{d}\rho  \, \chi_{\Lambda}^2(\rho) \, \frac{\rho}{(\alpha^2+\rho)} \int_0^{\pi}  \mathrm{d}\theta\, \sin\theta\,   (\cos\theta)^{2}\cos(2\rho y\cos\theta ).
\end{multline*}
A further change of variables $\tau = \cos\theta$ gives
\begin{align*}
&-\frac{1}{\pi}\int_0^{+\infty}\mathrm{d}\rho  \, \, \frac{\chi_{\Lambda}^2(\rho) \rho}{(\alpha^2+\rho)} \int_{-1}^{1}  \mathrm{d}\tau\; \tau^{2} \cos (2\rho y\tau) \\
&\quad= -\frac{1}{\pi}\int_0^{+\infty}\mathrm{d}\rho  \,  \, \frac{\chi_{\Lambda}^2(\rho)\rho}{(\alpha^2+\rho)} \left\{ \frac{-4 \sin(2\rho y) + 2(2 \rho y)^2 \sin(2 \rho y) + 8 y\rho \cos(2 \rho y)}{(2 \rho y)^3} \right\}.
\end{align*}
Changing again variables $\sigma = \rho y$ we have, recalling that $y = L \alpha^{-1}$,
\begin{equation*}
J = -\frac{1}{\pi y}\int_0^{+\infty}\mathrm{d}\sigma  \,  \frac{ \chi_{\Lambda}^2\left( \frac{\sigma}{y}\right)\sigma}{(y \alpha^2+\sigma)} \left(  \sin(2 \sigma) +\frac{\cos(2\sigma)}{\sigma} - \frac{\sin(2\sigma)}{2 \sigma^2}\right) = O\left( \frac{\alpha}{L}\right).
\end{equation*}
From this, using that
\begin{equation}
\left\langle x u_{\alpha}\,\bigg|\, \frac{(h_{\alpha}-e_{\alpha})^2}{(h_{\alpha}-e_{\alpha}+|k|)}\,\bigg|\, x u_{\alpha} \right\rangle = \alpha^2  \left\langle x u_{1}\,\bigg|\, \frac{(h_{1}-e_{1})^2}{(\alpha^2(h_{1}-e_{1})+|k|)}\,\bigg|\, x u_{1} \right\rangle,
\end{equation}
and the spectral theorem we conclude the proof.
\end{proof}

In the next lemma we show by complex line integration techniques, inspired by the original work of Casimir and Polder \cite{cp}, the main integral term gives the fundamental cancellation of the van der Waals term and produces the new leading term. 

\begin{proposition}\label{prop:cancellation_VdW}
For any $L>1$,
\begin{equation}
 \alpha \int_{\mathbb{R}^3} \mathrm{d} k \; f_y(k) \, \left\langle x u_{\alpha}\,\bigg|\, \frac{(h_{\alpha}-e_{\alpha})|k|}{3(h_{\alpha}-e_{\alpha}+|k|)}\,\bigg|\, x u_{\alpha} \right\rangle =\frac{\alpha^2}{L^3} - \aleph_{\alpha,L}\frac{\alpha}{L^4} + O\Big( \frac{\alpha^3}{L^2}\log(\alpha^{-1})\Big),
\end{equation}
where $\aleph_{\alpha,L}$ is defined in \eqref{eq:aleph_arctan_def}.
\end{proposition}
\begin{proof}
We write explicitly the expression of the integral and denote it by $I$,
\begin{align*}
I:=  \frac{\alpha}{4\pi^2} \int_{\mathbb{R}^3} \mathrm{d} k \; \frac{\chi^2_{\Lambda}(k)}{|k|} (-1-\hat{k}_1^2+\hat{k}_2^2+\hat{k}_3^2) \cos(2k_1y)\left\langle x u_{\alpha}\,\bigg|\, \frac{(h_{\alpha}-e_{\alpha})|k|}{3(h_{\alpha}-e_{\alpha}+|k|)}\,\bigg|\, x u_{\alpha} \right\rangle. 
\end{align*}
Let us pass to spherical coordinates $(\rho, \varphi, \theta) \in (0,+\infty) \times (0,2\pi) \times (0,\pi)$ so that $\rho = |k|$, $k_1 = \rho \cos(\theta)$ and $(-1-\hat{k}_1^2 + \hat{k}_2^2+ \hat{k}_3^2) = - 2 \cos^2\theta$,
\begin{align*}
-\frac{\alpha}{ 2\pi^2} \int_0^{+\infty}\int_{0}^{2\pi}\int_0^{\pi} \mathrm{d}\rho \,\mathrm{d}\varphi\,  \mathrm{d}\theta\, \rho^2  \sin\theta\, \chi_{\Lambda}^2(\rho)  \cos^2\theta \cos(2\rho y\cos\theta ) \left\langle x u_{\alpha}\,\bigg|\, \frac{(h_{\alpha}-e_{\alpha})}{3(h_{\alpha}-e_{\alpha}+\rho)}\,\bigg|\, x u_{\alpha} \right\rangle. 
\end{align*}
By an explicit calculation, the integration in 
$\varphi$ gives only a $2 \pi$ factor and the one in the $\theta$ variable: 
\begin{align*}
\int_0^{\pi} \, \mathrm{d}\theta \; \sin \theta \cos^2\theta \cos(2\rho y \cos\theta) &= \frac{\sin(2 \rho y)}{\rho y} + \frac{\cos(2 \rho y)}{\rho^2 y^2} - \frac{\sin(2\rho y)}{2 \rho^3 y^3}  \\
&= \left(\frac{-i}{2 \rho y} + \frac{1}{2 \rho^2 y^2}  + \frac{i}{4 \rho^3 y^3}  \right) e^{2 i  \rho y }  + \text{h.c.},
\end{align*}
where in the last line we used Euler formulas for sine and cosine. 
Plugging in the original calculation and making explicit the dependence on $\alpha$ we have 
\begin{align*}
I=\frac{\alpha}{ 6\pi} \int_0^{+\infty}\mathrm{d}\rho \;\rho^2  \, \chi_{\Lambda}^2(\rho)\,  \left\| \, \frac{(h_{1}-e_{1})^{1/2}}{(\alpha^2(h_{1}-e_{1})+\rho)^{1/2}}\, x u_{1} \right\|^2  \left\{\left(\frac{i}{ \rho y} - \frac{1}{ \rho^2 y^2}  - \frac{i}{2 \rho^3 y^3}  \right) e^{2 i  \rho y }  + \text{h.c.} \right\}. 
\end{align*}

Let us define the class of integrals below, for a measurable set $B \subseteq \mathbb{C}$:
\begin{align*}
I_B &:=  \int_B\mathrm{d}z \;g(z),\\
g(z) &:= \frac{\alpha}{ 6\pi} z^2  \, \chi_{\Lambda}^2(z)\,  \left\| \, \frac{(h_{1}-e_{1})^{1/2}}{(\alpha^2(h_{1}-e_{1})+z)^{1/2}}\, x u_{1} \right\|^2 \left(\frac{i}{ z y} - \frac{1}{ z^2 y^2}  - \frac{i}{2 z^3 y^3}  \right) e^{2 i  z y }. 
\end{align*}
Here $\chi_{\Lambda}(z) \equiv \chi_{\Lambda}(|z|), z \in \mathbb{C}$, denotes the complex extension of the frequencies cut-off defined only for real arguments.
Thanks to this notation, we can rewrite the integral over the half-line as a limit introducing a parameter $\eps \in (0,1)$: 
\begin{equation}
I = \lim_{\eps \rightarrow 0} (I_{(\eps, \eps^{-1})}  +\overline{I_{(\eps, \eps^{-1})}}).
\end{equation}

For both the integrals, their integrands are analytic in $\{z \in \mathbb{C}\,|\, \mathrm{Re} z >0\}\setminus B_{\eps}(0)$. Then, integrating over any closed path in that domain gives zero as result. Let us interpret the interval $(\eps, \eps^{-1})$ as part of two closed paths:
\begin{align*}
\gamma^+_{\eps, \eps^{-1}} &:= -\gamma^+_{\text{int}} \cup (\eps, \eps^{-1}) \cup \gamma^+_{\text{ext}} \cup i(\eps^{-1}, \eps)\\
\gamma^-_{\eps, \eps^{-1}} &:= \gamma^-_{\text{int}} \cup (\eps, \eps^{-1}) \cup -\gamma^-_{\text{ext}} \cup -i(\eps^{-1},\eps)
\end{align*}
where 
\begin{align*}
\gamma_{\text{int}}^{+} &= \Big\{z = \eps e^{i\theta}\,\Big|\, \theta \in  \Big(0,  \frac{\pi}{2}\Big)\Big\}, \qquad \gamma _{\text{ext}}^{+} = \Big\{z = \eps^{-1} e^{i\theta}\,\Big|\, \theta \in  \Big(0, \frac{\pi}{2}\Big)\Big\}, \\
\gamma_{\text{int}}^{-} &= \Big\{z = \eps e^{i\theta}\,\Big|\, \theta \in  \Big(-\frac{\pi}{2}, 0\Big)\Big\}, \qquad \gamma _{\text{ext}}^{-} = \Big\{z = \eps^{-1} e^{i\theta}\,\Big|\, \theta \in  \Big(-\frac{\pi}{2}, 0\Big)\Big\}.
\end{align*}

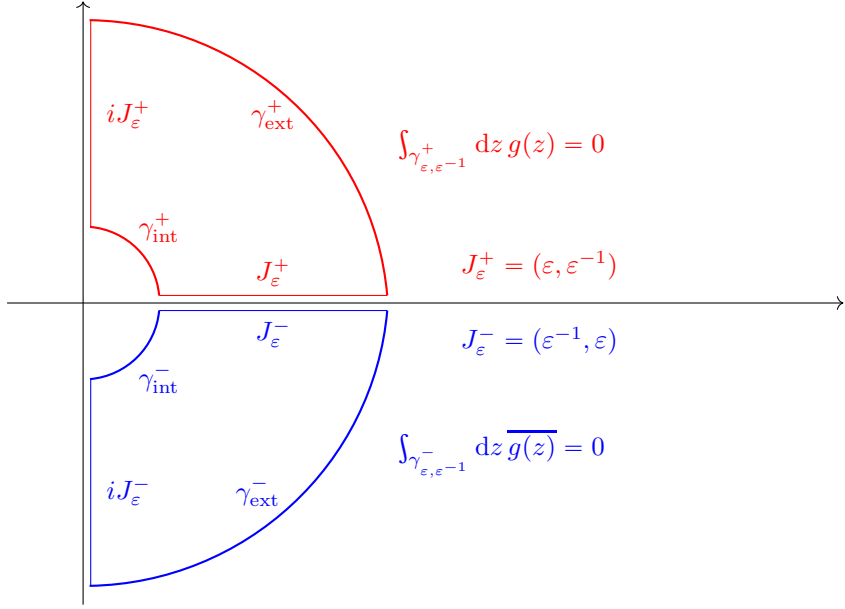
\begin{figure}
\caption{Complex line integrals}
\begin{center}
\begin{tikzpicture}
\draw[->] (-1,0) to (10,0);
\draw[->] (0,-4) to (0,4);
\draw[thick, red] (1,0.1) arc (5:85:1);
\draw[red] (1,0.1) to (4, 0.1);
\draw[red] (0.1,1) to (0.1, 3.75);
\draw[thick, red] (4,0.1) arc (5:89:4);
\draw[thick, blue] (1,-0.1) arc (-5:-85:1);
\draw[blue] (1,-0.1) to (4, -0.1);
\draw[blue] (0.1,-1) to (0.1, -3.75);
\draw[thick, blue] (4,-0.1) arc (-5:-89:4);
\node at (2.5, 0.4) {$\textcolor{red}{J_{\eps}^+}$};
\node at (2.5, -0.4) {$\textcolor{blue}{J^-_{\eps}}$};
\node at (1, 1) {$\textcolor{red}{\gamma^+_{\text{int}}}$};
\node at (1, -1) {$\textcolor{blue}{\gamma^-_{\text{int}}}$};
\node at (0.6, 2.5) {$\textcolor{red}{iJ^+_{\eps}}$};
\node at (0.6, -2.5) {$\textcolor{blue}{iJ^-_{\eps}}$};
\node at (2.5, 2.5) {$\textcolor{red}{\gamma_{\text{ext}}^+}$};
\node at (2.3, -2.5) {$\textcolor{blue}{\gamma_{\text{ext}}^-}$};
\node at (5.5,2) {  $\textcolor{red}{\int_{\gamma^+_{\eps,\eps^{-1}}} \mathrm{d} z \, g(z) =0}$};
\node at (5.5,-2) {  $\textcolor{blue}{\int_{\gamma^-_{\eps,\eps^{-1}}} \mathrm{d} z \, \overline{g(z)} =0}$};
\node at (6,0.5) {$\textcolor{red}{J_{\eps}^+ = (\eps,\eps^{-1})}$};
\node at (6,-0.5) {$\textcolor{blue}{J_{\eps}^- = (\eps^{-1},\eps)}$};
\end{tikzpicture}
\end{center}
\end{figure}
From the picture we see that 
\begin{align*}
I_{(\eps, \eps^{-1})} &= \int_{(\eps, \eps^{-1})} \mathrm{d}z\, g(z) = - \left[\int_{-\gamma^+_{\text{int}}} + \int_{i(\eps^{-1}, \eps) } + \int_{\gamma^+_{\text{ext}} } \right]\mathrm{d}z\, g(z),\\
\overline{I_{(\eps, \eps^{-1})}} &= \int_{(\eps, \eps^{-1})} \mathrm{d}z\, \overline{g(z)} = -\left[\int_{\gamma^-_{\text{int}}} + \int_{-i(\eps^{-1}, \eps) } + \int_{-\gamma^-_{\text{ext}} } \right] \mathrm{d}z \, \overline{g(z)}.
\end{align*}

We can immediately observe that the ext terms disappear in the limit of $\eps \rightarrow 0$. Indeed $g(\varepsilon^{-1}e^{i\theta}) = 0 $, for sufficiently large $\varepsilon^{-1}$ (bigger than $\Lambda$), thanks to the presence of the cut-off $\chi_{\Lambda}$:
\begin{equation}\label{eq:ext_int}
\lim_{\eps \rightarrow 0 } \int_{\gamma^{\pm}_{\text{ext}}} \mathrm{d} z\; g(z)^{\#} = 0,
\end{equation}
with $g^{\#}$ covering both the cases $g$ and $\bar{g}$.

So it remains to analyze the int and imaginary terms. 
Changing variables to $z = \eps e^{i \theta}$ in the integral over $\gamma_{\text{int}}^+$, we have
\begin{align*}
\int_{\gamma_{\text{int}}^+} \mathrm{d} z\; g(z) = \frac{\alpha}{ 6\pi}\int_0^{\frac{\pi}{2}} i \eps e^{i\theta} \mathrm{d}\theta&\;  \eps^2 e^{2 i\theta} \, \chi_{\Lambda}^2(\eps)\,  \left\| \, \frac{(h_{1}-e_{1})^{1/2}}{(\alpha^2(h_{1}-e_{1})+ \eps e^{i \theta})^{1/2}}\, x u_{1} \right\|^2\\
&\;\times \left(\frac{i}{ \eps e^{i \theta} y} - \frac{1}{ \eps^2e^{2i\theta} y^2}  - \frac{i}{2 \eps^3 e^{3i\theta} y^3}  \right) e^{2 i  \eps e^{i \theta} y }.
\end{align*}
By the bounded convergence theorem we can switch the limit with the integral and obtain, having in mind that $\|xu_1\|^2 = 12$ and recalling that $y = L\alpha^{-1},$
\begin{align*}
-\lim_{\eps \rightarrow 0} \int_{-\gamma_{\text{int}}^+}\mathrm{d} z \; g(z) &= \frac{\alpha^{-1}}{ 6\pi}\int_0^{\frac{\pi}{2}}   e^{3i\theta} \mathrm{d}\theta\;   \,  \left\|  x u_{1} \right\|^2 \frac{1}{2 e^{3i\theta} y^3}\\
&=  \frac{\alpha^{-1}}{6 \pi} \frac{\pi}{2} \frac{12}{2 y^3} = \frac{\alpha^2}{2 L^3}.
\end{align*}
A totally analogous calculation yields the same result for $\gamma^-_{\text{int}}$, so that 
\begin{equation}\label{eq:resultsinterior}
- \lim_{\eps \rightarrow 0} \left\{\int_{-\gamma_{\text{int}}^+}\mathrm{d} z \; g(z) +\int_{\gamma_{\text{int}}^-}\mathrm{d} z \; \overline{g(z)}  \right\} = \frac{\alpha^2}{L^3}.
\end{equation}

We show how the term of order $\frac{\alpha}{L^4}$ comes from the integration on the imaginary axis. For the path in $i(\eps, \eps^{-1})$ we change variable setting $w = -i z$ and denoting by $b_y(w) := \left(\frac{1}{ w y} + \frac{1}{ w^2 y^2}  + \frac{1}{2 w^3 y^3}  \right) e^{-2 w y }$:
\begin{align*}
&\int_{i(\eps^{-1}, \eps) } \mathrm{d}z\, g(z) \\
&= \frac{\alpha}{ 6\pi} \int_{\eps}^{\eps^{-1}} i \mathrm{d}w\; w^2  \, \chi_{\Lambda}^2(w)\,  \left\| \, \frac{(h_{1}-e_{1})^{1/2}}{(\alpha^2(h_{1}-e_{1})+i w)^{1/2}}\, x u_{1} \right\|^2 b_y(w) \\
&= \frac{\alpha}{ 6\pi} \int_{\eps}^{\eps^{-1}}  \mathrm{d}w\; w^2  \, \chi_{\Lambda}^2(w)\,  \left\langle x u_1\,\bigg|\, \frac{(h_{1}-e_{1})}{ (-i\alpha^2(h_{1}-e_{1})+w)}\,\bigg|\, x u_{1}\right\rangle b_y(w), 
\end{align*}
while for the path $i(\eps, \eps^{-1})$ we change variable setting $w = i z$, obtaining
\begin{align*}
\int_{-i(\eps^{-1}, \eps) } \mathrm{d} z \; \overline{g(z)}= \frac{\alpha}{ 6\pi} \int_{\eps}^{\eps^{-1}}  \mathrm{d}w\; w^2  \, \chi_{\Lambda}^2(w)\,  \left\langle x u_1\,\bigg|\, \frac{(h_{1}-e_{1})}{ (i\alpha^2(h_{1}-e_{1})+w)}\,\bigg|\, x u_{1}\right\rangle b_y(w).
\end{align*}
Using that 
\begin{equation}
\left\langle x u_1\,\bigg|\,\frac{(h_{1}-e_{1})}{ (-i\alpha^2(h_{1}-e_{1})+w)} +  \frac{(h_{1}-e_{1})}{ (i\alpha^2(h_{1}-e_{1})+w)}\,\bigg|\, x u_{1}\right\rangle = \left\langle x u_1\,\bigg|\,\frac{2 w (h_{1}-e_{1})}{ (\alpha^4(h_{1}-e_{1})^2+w^2)} \,\bigg|\, x u_{1}\right\rangle 
\end{equation}
we sum the two contributions to finally obtain
\begin{multline*}
- \int_{i(\eps^{-1}, \eps) } \mathrm{d}z\, g(z)- \int_{-i(\eps^{-1}, \eps) } \mathrm{d} z \; \overline{g(z)}  \\
= -\frac{\alpha}{ 6 \pi y^3}  \int_{\eps}^{\eps^{-1}}  \mathrm{d}w\;  \, \chi_{\Lambda}^2(w)\, \left\|\frac{  (h_{1}-e_{1})^{1/2}x u_{1}}{ (\alpha^4(h_{1}-e_{1})^2+w^2)^{1/2}} \, \right\|^2  2 w^3 y^3 b_y(w).
\end{multline*}
By a further change of variable $v = wy = w L\alpha^{-1}$,
\begin{multline*}
- \int_{i(\eps^{-1}, \eps) } \mathrm{d}z\, g(z)- \int_{-i(\eps^{-1}, \eps) } \mathrm{d} z \; \overline{g(z)} =\\
=-\frac{\alpha^5}{ 6 \pi L^4}  \int_{y \eps}^{y \eps^{-1}}  \mathrm{d}v\;  \, \chi_{\Lambda}^2\left(\frac{\alpha v}{L}\right)\, \left\|\frac{  (h_{1}-e_{1})^{1/2}x u_{1}}{ (\alpha^4(h_{1}-e_{1})^2+L^{-2}\alpha^2 v^2)^{1/2}} \, \right\|^2  (2 v^2 + 2 v  + 1  ) e^{-2 v }  \\
\xrightarrow[\eps \rightarrow 0]{} -\frac{\alpha^3}{ 6 \pi L^4}  \int_{0}^{+\infty}  \mathrm{d}v\;  \, \chi_{\Lambda}^2\left(\frac{\alpha v}{L}\right)\, \left\|\frac{  (h_{1}-e_{1})^{1/2}x u_{1}}{ (\alpha^2(h_{1}-e_{1})^2+\frac{v^2}{L^2})^{1/2}} \, \right\|^2  \left(2 v^2 + 2 v  + 1  \right) e^{-2 v }.
\end{multline*}
Let us split the region of integration in two parts in order to estimate the last integral: $(0, 1)\cup (1, +\infty)$.

\begin{itemize}
\item $v \in (1, +\infty)$: the following estimates holds
\begin{multline*}
\frac{\alpha^3}{6 \pi L^4}\bigg|\int_{1}^{+\infty}  \mathrm{d}v\;  \, \chi_{\Lambda}^2\left(\frac{\alpha v}{L}\right)\, \bigg\|\frac{  (h_{1}-e_{1})^{1/2}x u_{1}}{ (\alpha^2(h_{1}-e_{1})^2+\frac{v^2}{L^2})^{1/2}} \, \bigg\|^2  \left(2 v^2 + 2 v  + 1  \right) e^{-2 v } \bigg|\\
\leq  \frac{\alpha^3}{L^2} \left\|  (h_{1}-e_{1})^{1/2}x u_{1}  \right\|^2 \int_{1}^{+\infty}  \mathrm{d}v\;      e^{-2 v } = O\Big( \frac{\alpha^3}{L^2}\Big).
\end{multline*}

\item $v \in (0,1)$: the approximation of $\chi_{\Lambda}^2 (2v^2+2v+1)e^{-2v}$ by $1$ thanks to a Taylor expansion produces the following error
\begin{equation}
\Big|\chi_{\Lambda}^2 \big(\frac{\alpha v}{L}\big)(2v^2+2v+1)e^{-2v} -1 \Big|\leq C \Big( \frac{\alpha}{L} \|\nabla \chi_{\Lambda}\|_{\infty} +1 \Big)|v|,
\end{equation}
which, by the functional calculus, implies
\begin{equation*}
\frac{\alpha^3}{6 \pi L^4}\int_{0}^{1}  \mathrm{d}v \, \left\|\frac{  (h_{1}-e_{1})^{1/2}x u_{1}}{ (\alpha^2(h_{1}-e_{1})^2+\frac{v^2}{L^2})^{1/2}} \, \right\|^2  \left|\chi_{\Lambda}^2\left(\frac{\alpha v}{L}\right)\left(2 v^2 + 2 v  + 1  \right) e^{-2 v }-1 \right| = O\Big( \frac{\alpha^3}{L^2} \log(\alpha^{-1})\Big). 
\end{equation*}
Therefore we pass to estimate the integral below, where, again by the functional calculus, we can write
\begin{equation*}
-\frac{\alpha^3}{6 \pi L^4} \int_{0}^{1}  \mathrm{d}v\;   \left\|\frac{  (h_{1}-e_{1})^{1/2}x u_{1}}{ (\alpha^2(h_{1}-e_{1})^2+\frac{v^2}{L^2})^{1/2}} \, \right\|^2= -\frac{\alpha}{6 \pi L^4} \left\langle\alpha L\arctan\left(\frac{1}{\alpha L (h_1 - e_1)}\right) \right\rangle_{x u_1}.
\end{equation*}
Introducing the quantity
\begin{equation}\label{eq:aleph_arctan_def}
\aleph_{\alpha,L} :=  \frac{1}{6 \pi } \left\langle\alpha L\arctan\left(\frac{1}{\alpha L (h_1 - e_1)}\right) \right\rangle_{x u_1}, 
\end{equation}
we can finally state that
\begin{equation}\label{eq:imaginary_int}
\lim_{\varepsilon \rightarrow 0}\left\{- \int_{i(\eps^{-1}, \eps) } \mathrm{d}z\, g(z)- \int_{-i(\eps^{-1}, \eps) } \mathrm{d} z \; \overline{g(z)}\right\} = -\frac{\alpha}{L^4} \aleph_{\alpha,L} + O\Big( \frac{\alpha^3}{L^2}\log(\alpha^{-1})\Big). 
\end{equation}
\end{itemize}
Collecting the estimates \eqref{eq:ext_int}, \eqref{eq:resultsinterior} and \eqref{eq:imaginary_int}, we conclude the proof of Proposition \ref{prop:cancellation_VdW}.
\end{proof}

\section*{Appendix 3: Derivation of the model: quantization on half space}\label{app:quantization}

For simplicity we consider the half space 
\[\mathbb{R}^3_+ := \{ x=(x_1,x_2,x_3) \in \mathbb{R}^3\,|\, x_1 >0\},\]
with the surface of the conductor being 
$\Sigma_0 = \{(0,x_2,x_3) \in \mathbb{R}^3\}$, and we obtain the general result by translation and reflection.
We denote by $\mathcal E(x,t) = (\mathcal E^{(j)}(x,t))_{j=1}^3,$ and $\mathcal{B}(x,t) =(\mathcal{B}^{(j)}(x,t))_{j=1}^3$, $x \in \mathbb{R}^3_+$, the components of the classical electric and magnetic fields $\mathcal{E}, \mathcal{B} \in \mathbb{R}^3$, respectively.

The standard boundary conditions for $(\mathcal{E},\mathcal{B})$ in the presence of a grounded, perfect conductor wall which can be found, for example, in formula (13.106) from \cite{spohnbook}, are: 
\begin{equation}
\hat{n}(x) \times \mathcal E(x) = 0, \qquad \hat{n}(x) \cdot \mathcal B(x) = 0, \qquad \text{for any } x \in \Sigma_0,
\end{equation}
where $\hat{n}$ denotes the outward normal versor to the surface of the wall, in our case $\hat{n}(x) = (1, 0, 0)$. This implies that the conditions can be rewritten as
\begin{equation}
\mathcal E^{(2)}(0,x_2,x_3) = 0 = \mathcal E^{(3)}(0,x_2,x_3),\qquad \mathcal{B}^{(1)}(0,x_2,x_3) = 0, \qquad (x_2,x_3) \in \mathbb{R}^2.
\end{equation}
We start from observing that, in the classical setting, the electric field function has to be a solution of the wave equation in the half space with constraints given by the aforementioned boundary conditions for the conductor surface:
\begin{equation}
\begin{cases}
\partial_t^2 \mathcal{E}^{(j)}(x,t) = - \Delta_x \mathcal{E}^{(j)}(x,t), \quad & x_1 >0,\\
\mathcal{E}^{(j)}(0,x_2,x_3, t) = 0,  \quad & j=2,3.
\end{cases}
\end{equation}
We introduce a new electric field on the full space by an odd reflection
\begin{equation}
\widetilde{\mathcal{E}}^{(j)}(x,t) := 
\begin{cases}
\mathcal{E}^{(j)}(x,t), \quad & \text{if } x_1 \geq 0,\\
-\mathcal{E}^{(j)}(-x_1,x_2,x_3,t), \quad & \text{if } x_1<0. 
\end{cases}
\end{equation}
The field is assumed to be real and its expansion in Fourier modes as solution of the wave equation  has the standard expression
\begin{align}\label{elecfield:full}
\widetilde{\mathcal{E}}^{(j)}(x,t) &= \frac{1}{(2\pi)^{3/2}} \int_{\mathbb{R}^3} \diff k \, (\beta^{(j)}_+(k) e^{i(kx-\omega t)} + \beta^{(j)}_-(k) e^{-i(kx-\omega t)}), \\
 \beta_{+}^{(j)}(k) &:= \mathscr{F}[\widetilde{\mathcal{E}}^{(j)}(\cdot,0)](k), \qquad \beta_{-}^{(j)}(k) =  \overline{\beta_{+}^{(j)}(k)},
\end{align}
where we denoted by $\mathscr{F}$ the Fourier transform in $\mathbb{R}^3$.
Since $\widetilde{\mathcal{E}}^{(j)}$ is odd in $x_1$, its Fourier transform is odd in $k_1$ and this implies the following relations for the coefficients:
\begin{equation}
\beta_{\pm}^{(j)}(-k_1,k_2,k_3) = -\beta_{\pm}^{(j)}(k), \qquad j=2,3,
\end{equation}
which gives back, using an odd reflection, an expansion for the electric field \eqref{elecfield:full} in terms of sines in the $x_1$ direction: for $j=2,3$,
\begin{align*}
\widetilde{\mathcal{E}}^{(j)}(x,t)
&= \frac{2}{(2\pi)^{3/2}} \int_{\mathbb{R}^3_+} \diff k \, \sin(k_1 x_1)(i \beta^{(j)}_+(k) e^{i(k_2 x_2 + k_3 x_3-\omega t)} - i \beta^{(j)}_-(k) e^{-i(k_2 x_2 + k_3 x_3-\omega t)})  \\
&= \frac{2}{(2\pi)^{3/2}} \sum_{\gamma = 1,2}\int_{\mathbb{R}^3_+} \diff k \, \sin(k_1 x_1)\mathbf{e}^{(j)}_{\gamma}(k)(i \beta_{+,\gamma}(k)  e^{i(k_2 x_2 + k_3 x_3-\omega t)} + h.c.),
\end{align*}
where we projected the Fourier coefficients on the polarization vectors $\{\mathbf{e}_{\gamma}(k)\}_{\gamma =1,2}$
\begin{equation}
\vec{\beta}_{\pm}(k) = \sum_{\gamma =1,2} \beta_{\pm,\gamma}(k) \mathbf{e}_{\gamma}(k), \qquad \vec{\beta}_{\pm}(k) = (\beta^{(1)}_{\pm}, \beta^{(2)}_{\pm}, \beta_{\pm}^{(3)}),
\end{equation}
and assumed to work in Coulomb gauge. 

Recalling the Maxwell equations in the vacuum 
\begin{align*}
&\nabla \cdot \widetilde{\mathcal{E}} = 0, \qquad  \nabla \times \widetilde{\mathcal{B}}=\der_t \widetilde{\mathcal{E}},  \\
&\nabla \cdot \widetilde{\mathcal{B}} = 0, \qquad \nabla \times \widetilde{\mathcal{E}}=-\der_t \widetilde{\mathcal{B}}, 
\end{align*}
we can recover the expressions of the components of $\widetilde{\mathcal{B}}$ and of $\widetilde{\mathcal{E}}^{(1)}$.

We introduce the classical vector potential $\mathcal{A}$ on $\mathbb{R}^3_+$ and its extension $\widetilde{\mathcal{A}}$ on $\mathbb{R}^3$, even for the first component and odd for the remaining ones. By the equation $\widetilde{\mathcal{B}} = \nabla \times \widetilde{\mathcal{A}}$ we recover the expression of $\widetilde{\mathcal{A}}$ as well. Collecting the previous formulas for the expansions we finally obtain, for all the fields,
\begin{align}
\widetilde{\mathcal{E}}(x,t) &= \frac{1}{(2\pi)^{3/2}} \sum_{\gamma = 1,2}\int_{\mathbb{R}^3_+} \diff k \;\beta_{+,\gamma}(k) \,\mathbf{b}(k) + h.c.,\\
\widetilde{\mathcal{B}}(x,t) &= \frac{1}{(2\pi)^{3/2}} \sum_{\gamma = 1,2}\int_{\mathbb{R}^3_+} \diff k \,\beta_{+,\gamma}(k) \, k \times \mathbf{b}(k)+ h.c., \\
\widetilde{\mathcal{A}}(x,t) &= \frac{1}{(2\pi)^{3/2}} \sum_{\gamma = 1,2}\int_{\mathbb{R}^3_+} \frac{\diff k}{\omega(k)} \beta_{+,\gamma}(k) \,\mathbf{b}(k) + h.c.,
\end{align}
where 
\begin{equation}
\mathbf{b}(k) := \left(\begin{array}{c}
2\cos(k_1 x_1) \mathbf{e}^{(1)}_{\gamma}(k) e^{i(k_2 x_2 + k_3 x_3-\omega t)}\\
2i\sin (k_1 x_1) \mathbf{e}^{(2)}_{\gamma}(k)  e^{i(k_2 x_2 + k_3 x_3-\omega t)} \\
2i\sin (k_1 x_1) \mathbf{e}^{(3)}_{\gamma}(k) e^{i(k_2 x_2 + k_3 x_3-\omega t)} 
\end{array}\right).
\end{equation}
We introduce the rescaled Fourier coefficients $\{\alpha_{\pm,\gamma}\}_{\gamma=1,2}$ by
\begin{equation}\label{eq:symmalpha}
\beta_{\pm,\gamma}(k) = (2\pi)^{3/2} \frac{\omega^{1/2} (k)}{2 \pi} \alpha_{\pm,\gamma}(k).
\end{equation}
We further introduce a cut-off $\chi_{\Lambda} \in C_0^{\infty}(\mathbb{R}^3)$ for the momenta (see the construction of the Abraham model \cite[Chapter 2.4]{spohnbook}) and we derive the expansion expression for the original fields $(\mathcal{E},\mathcal{B}, \mathcal{A})$, for $x \in \mathbb{R}^3_+$,
\begin{equation}
\mathcal{E}(x,t) = \widetilde{\mathcal E}(x,t)|_{x_1>0}, \qquad \mathcal B(x,t) = \widetilde{\mathcal B}(x,t)|_{x_1>0}, \qquad \mathcal A(x,t) = \widetilde{\mathcal A}(x,t)|_{x_1>0},
\end{equation}  
where
\begin{equation}
\widetilde{\mathcal{A}}(x,t) = \sum_{\gamma = 1,2}\int_{\mathbb{R}^3_+} \diff k \, \frac{\chi_{\Lambda}(k)}{ 2 \pi |k|^{1/2}} \alpha_{+,\gamma}(k)  \left(\begin{array}{c}
2\cos(k_1 x_1) \textbf{e}^{(1)}_{\gamma}(k) e^{i(k_2 x_2 + k_3 x_3-\omega t)}\\
2i\sin (k_1 x_1) \textbf{e}^{(2)}_{\gamma}(k)  e^{i(k_2 x_2 + k_3 x_3-\omega t)} \\
2i\sin (k_1 x_1) \textbf{e}^{(3)}_{\gamma}(k)  e^{i(k_2 x_2 + k_3 x_3-\omega t)} 
\end{array}\right) + h.c.
\end{equation}
We want to derive the Fourier modes expansion for the electromagnetic energy too. By the usual definition, this time adapted to the half space,
\begin{align*}
h_f &:= \frac{1}{8\pi}\int_{\mathbb{R}^3_+} \diff x\, (|\mathcal{E}(x,t)|^2 + |\mathcal{B}(x,t)|^2) \\
&= \frac{1}{8\pi} \sum_{j=1}^3 \int_{\mathbb{R}^3_+} \diff x\, (|\mathcal{E}^{(j)}(x,t)|^2 + |\mathcal{B}^{(j)}(x,t)|^2).
\end{align*}
Comparing the integral with the odd extensions for $j=2,3$ and with the even extension for $j=1$, we can write
\begin{align*}
h_f &= \frac{1}{16\pi} \sum_{j=1}^3 \int_{\mathbb{R}^3} \diff x\, (|\widetilde{\mathcal{E}}^{(j)}(x,t)|^2 + |\widetilde{\mathcal{B}}^{(j)}(x,t)|^2)  \\
&= \frac{1}{2} \sum_{\gamma=1,2} \int_{\mathbb{R}^3} \diff k\, |k|\alpha_{-,\gamma}(k) \alpha_{+,\gamma}(k) = \sum_{\gamma=1,2}\int_{\mathbb{R}^3_+}  \diff k\, |k|\alpha_{-,\gamma}(k) \alpha_{+,\gamma}(k),
\end{align*}
where for the second equality we used the usual expression for the electromagnetic energy in the full space and in the third equality we used \eqref{eq:symmalpha} and the symmetry properties of the $\beta$'s to change the domain of integration.

By Wick quantization techniques for polynomial symbols (see \cite{an} for details) we can define $(E, B, A)$ being the associated quantum field versions of the electromagnetic operators $(\mathcal{E}, \mathcal{B}, \mathcal{A})$, respectively. The theory results into the intuitive quantization rules
\begin{equation}
\alpha_{+,\gamma}(k) \quad \rightarrow \quad a_{\gamma}^{\dagger}(k), \qquad \alpha_{-,\gamma}(k) \quad \rightarrow \quad a_{\gamma}^{}(k),
\end{equation}
substitution that for a polynomial symbol $p(\alpha_{+},\alpha_-)$ we denote as $(p(\alpha_+,\alpha_-))^{\mathrm{Wick}}$. In this way we can write
\begin{equation}
A (x) := (\mathcal{A}(x,0))^{\mathrm{Wick}}, \qquad H_f^+ := (h_f)^{\mathrm{Wick}}, \qquad x \in \mathbb{R}^3_+
\end{equation}
which gives the same expression for the operators $A_y(x), H_f$ given in Section \ref{sec:interaction_model} by a translation and a reflection in the $x_1$ variable for $A(x)$.

\bibliographystyle{plain} 
\bibliography{CP}

\end{document}